\documentclass{bmcart}

\usepackage[utf8]{inputenc} 
\usepackage{bm}
\usepackage{amsthm,amsmath}
\usepackage{amssymb}
\usepackage{graphicx}
\usepackage{subfigure}
\usepackage{cite}
\usepackage{algorithmic}
\usepackage{algorithm}
\usepackage{multirow}
\usepackage{color}

\providecommand{\tabularnewline}{\\}

\theoremstyle{plain}
\newtheorem{thm}{\protect\theoremname}
\theoremstyle{remark}
\newtheorem{rem}[thm]{\protect\remarkname}
\theoremstyle{definition}
\newtheorem{example}[thm]{\protect\examplename}
\theoremstyle{plain}
\newtheorem{lem}[thm]{\protect\lemmaname}

\usepackage{bm}
\usepackage{amsthm,amsmath}
\usepackage{amssymb}
\usepackage{graphicx}
\usepackage{subfigure}
\usepackage{cite}
\usepackage{algorithmic}
\usepackage{algorithm}

\makeatother

\providecommand{\examplename}{Example}
\providecommand{\lemmaname}{Lemma}
\providecommand{\remarkname}{Remark}
\providecommand{\theoremname}{Theorem}

\startlocaldefs
\endlocaldefs

\makeatother

\begin{document}

\begin{frontmatter}

\begin{fmbox}
\dochead{Review}


\title{Energy-Efficient Transmission Strategies for CoMP Downlink –- Overview, Extension, and Numerical Comparison}


\author[
   addressref={aff1},                   
    noteref={n1},                    
   email={giang.nguyen@oulu.fi}   
]{\inits{1}\fnm{Kien-Giang} \snm{Nguyen}}
\author[
   addressref={aff1},
   noteref={n1},
   email={oskari.tervo@oulu.fi}
]{\inits{2}\fnm{Oskari} \snm{Tervo}}
\author[
   addressref={aff1},
   email={doanh.vu@oulu.fi}
]{\inits{3}\fnm{Quang-Doanh} \snm{Vu}}
\author[
   addressref={aff2},
   email={nam.tran@ucd.ie}
]{\inits{5}\fnm{Le-Nam} \snm{Tran}}
\author[
   addressref={aff1},
   email={markku.juntti@oulu.fi}
]{\inits{4}\fnm{Markku} \snm{Juntti}}

%

\address[id=aff1]{
  \orgname{Centre for Wireless Communications, University of Oulu, Oulu, Finland}, 
  \postcode{P.O.Box 4500, FI-90014}                                
  \city{Oulu},                              
  \cny{Finland}                                    
}
\address[id=aff2]{%
  \orgname{School of Electrical and Electronic Engineering, University College Dublin},
  \city{Dublin},
  \cny{Ireland}
}

\begin{artnotes}
\note[id=n1]{Equal contributor} 
\end{artnotes}



\end{fmbox}


\begin{abstractbox}

\begin{abstract}
This paper focuses on energy-efficient coordinated multi-point (CoMP) downlink in multi-antenna multi-cell wireless communications systems. We provide an overview of transmit beamforming designs for various energy efficiency (EE) metrics including maximizing the overall network EE, sum weighted EE and fairness EE. Generally, an EE optimization problem is a nonconvex program for which finding the globally optimal solutions requires high computational effort. Consequently, several low-complexity suboptimal approaches have been proposed. Here we sum up the main concepts of the recently proposed algorithms based on the state-of-the-art successive convex approximation (SCA) framework. Moreover, we discuss the application to the newly posted EE problems including new EE metrics and power consumption models. Furthermore, distributed implementation developed based on alternating direction method of multipliers (ADMM) for the provided solutions is also discussed. For the sake of completeness, we provide numerical comparison of the SCA based approaches and the conventional solutions developed based on parametric transformations (PTs). We also demonstrate the differences and roles of different EE objectives and power consumption models.
\end{abstract}




\end{abstractbox}
%

\end{frontmatter}




\section{Introduction}

Fifth generation (5G) wireless network visions foresee the challenges
of the data traffic demand caused by the upcoming explosive growth
of wireless devices and applications \cite{1000xdata}. The rapid
expansion of mobile networks is increasing the energy consumption
beyond sustainable limits. In the larger base stations (BSs), the most power-hungry components of the
multi-antenna transmitters are the transmit power amplifiers (PAs),
but the other circuits and components are also significant power consumers. In fact, they become even dominant in the smaller BSs, which are becoming more and more popular in the future dense networks. Nevertheless, this
causes problems in terms of electricity costs for operators and the
increase in greenhouse gas emission for the whole world \cite{FengJiang:13:aSurveyofEE,GYLiEESurveyTutorial}.
Consequently, energy efficiency (EE) has become an important design
target for wireless access networks.

In wireless communications, energy efficiency is generally defined as the ratio of the total reliably transmitted data to the total energy consumption \cite{Isheden:2012:WCOM}. In other words, it equals the achievable data rate in bits per second divided by the consumed power in Watts.  In either case, the basic unit of EE is bits per Joule (bits/J). It is worth mentioning that the classical transceiver optimization framework, on the other hand, typically focuses on maximizing the multiuser weighted sum rate or (area) spectral efficiency regardless of the proportionally rapid increase of total power consumed by the wireless network. The EE optimization deviates from this set-up by making a controlled trade-off between the supported rate and the consumed power
\cite{FengJiang:13:aSurveyofEE,GYLiEESurveyTutorial,zappone2015energy,Tervo-17,Oskari:EE-optimalBeamDesign:14:JSP}.

Variations of the EE objective have been proposed depending on the
system constraints and design targets. The basic alternatives include
\emph{network EE} (NEE), \emph{sum weighted EE} (SWEE) and \emph{fairness
EE} \cite{zappone2015energy}. While the first metric optimizes the
EE gain of the entire network, the others aim at satisfying the specific
EE requirements on individual base stations or users involved.

In the presence of multiuser interference, an EE maximization (EEmax)
problem belongs to a class of non-convex fractional programs for which
finding a globally optimal solution is challenging. However, an
optimal solution of the EEmax problem in multiuser multiple-input
single-output (MISO) downlink system has been provided in \cite{Oskari:EE-optimalBeamDesign:14:JSP}
using a branch-reduce-and-bound approach. Even though this approach guarantees finding the global optimum, it still requires very high
computational complexity. Therefore, low-complexity suboptimal designs
have attracted more attention for practical applications.

Common suboptimal approaches for EE designs have been developed
based on parametric transformation (PT) inspired by the fractional
structure of the EE objectives \cite{zappone2015energy,DerrickKwanNg:2012:JWCOM:EE_OFDM,She:2014:WSEE}.
However, such an approach leads to two-layer iterative procedures
\cite{She:2014:WSEE}, which often have high computational complexity
(as discussed in Section \ref{subsec:Conventional-Approaches}) and/or
are not suitable for distributed implementation. In addition, analyzing
the convergence of those methods has not been properly addressed \cite{Oskari:EE-optimalBeamDesign:14:JSP}.

Recently, novel algorithms have been developed based on the state-of-the-art
local optimization toolbox, namely successive convex approximation
(SCA) algorithm, which efficiently solves the EEmax problems; the
proposed framework is a one-loop iterative procedure which finds out
locally optimal solutions after a relatively small number of iterations,
and, thus, significantly reduces the complexity compared to the existing
PT approach \cite{Giang:15:JCOML}; the convergence of the SCA based
methods is provably guaranteed \cite{Giang:15:JCOML,Oskari:EE-optimalBeamDesign:14:JSP},
and the procedure is also well suited for implementation in a distributed
manner \cite{Giang:2017:WCOM}.

In this paper, we consider coordinated multi-point (CoMP) downlink
in multi-antenna multi-cell systems and focus on the applications
of the SCA approach on the EEmax problems arising in the wireless
access systems such as 4G and 5G cellular standards. The main contributions
of this paper can be summarized as follows:
\begin{itemize}
\item \emph{Overview}: We provide a summary of the basic concepts of the
SCA based algorithms, introduce some key transformations which turn
the EEmax problems into representations that successfully leverage
the principle of the SCA, revisit the problems of maximizing the NEE,
SWEE and maxminEE, and discuss how to arrive at efficient solutions.
We also discuss how to distributively implement the solutions.
\item \emph{Extension}: We discuss the recently proposed weighted product EE (WPEE) objective function and a general model of power consumption. We show how to adopt the proposed framework to the EEmax problems involved.
\item \emph{Numerical comparisons}: We make several numerical comparisons
on the algorithms. The most important one is the comparison between
the existing and the proposed approaches in terms of convergence speed
and average performances. Other evaluations have been made
to illustrate the roles and benefits of different EE objectives, and
the impact of different power consumption models on the EE performance.
\end{itemize}
An initial version of the paper was published in \cite{Nguyen:EuCNC}.
Herein we provide a more detailed and broader summary of the EE optimization
and discussion on the differences of the SCA and fractional programming
based approaches. We also extend the SCA framework to solve the problem
of WPEE maximization. We further present four different approximations
for the involved logarithmic functions, which enable second-order
programming formulations of the problems. Finally, we consider more
detailed power consumption models and provide a significantly more
extensive set of simulation results to evaluate different methods.

The rest of the paper is organized as follows. System model and several
energy efficiency measures are presented in Section \ref{sec:sysmodel}.
Centralized solutions and their distributed implementation are provided
in Section \ref{sec:centraldistributedsols}, followed by numerical
results in Section \ref{sec:Numerical-result}. Conclusion is provided
in Section \ref{sec:conclude}.

\emph{Notation}: Bold lower and upper case letters represent vectors
and matrices, respectively; calligraphic letters denote
sets; $\left|\cdot\right|$ represents the absolute value; $\left\Vert \cdot\right\Vert _{2}$
represents the $l_{2}$ norm; $\mathcal{CN}(0,a)$ denotes a zero
mean circularly symmetric complex Gaussian random variable with variance
$a$; $\mathbb{C}^{a\times b}$ represents the space of complex matrices
of dimensions given in superscript; $\mathrm{\Re}(\cdot)$ represents
real part of the argument; $\mathbb{E}\{\cdot\}$ denotes the expectation
operator. $\mathbf{a}^{T}$ and $\mathbf{a}^{H}$ stand for the transpose
and the Hermitian transpose of $\mathbf{a}$, respectively. $\left\langle \mathbf{a},\mathbf{b}\right\rangle $
denotes the inner product of vectors $\mathbf{a}$ and $\mathbf{b}$.
$\{\mathbf{a}_{b}\}_{b\in\mathcal{B}}$ refers to a composite vector containing all $\mathbf{a}_{b}$ where $b$ belongs to the set
$\mathcal{B}$. $\nabla_{{\bf x}}g({\bf x})$ represents
the partial derivative of function $g({\bf x})$ with respect to the
elements of ${\bf x}$. Other notations are defined at their first
appearance.

\section{System Model and Energy-Efficient Problem Formulations}

\label{sec:sysmodel}

\subsection{Channel and Signal Model}

 \begin{figure}
 \centering{}\includegraphics[width=0.8\columnwidth]{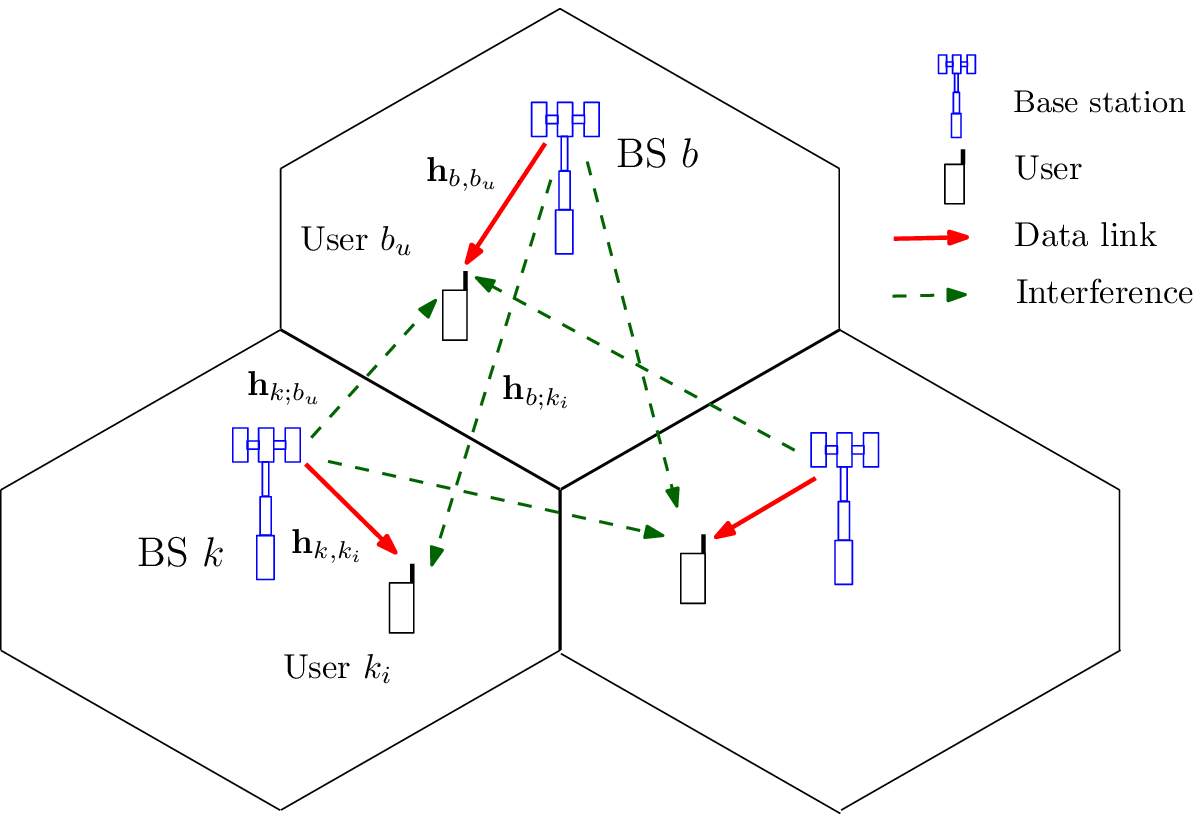}\caption{CoMP system model}\label{fig.sim}
 \end{figure}
We consider a downlink transmission in multi-cell multi-user multiple-input
single-output (MISO) system consisting of $B$ BSs,
each of which is equipped with $M$ antennas. There are $U$ single-antenna
users in each cell and a total of $UB$ users in the network\footnote{An equal number of users in each cell are assumed merely for lightening
up the notations.}. We assume that the BSs operate following the coordinated beamforming
mode, i.e., each BS only serves $U$ users in its own cell.\footnote{The algorithm frameworks provided in this paper can be straightforwardly
extended to the case of joint transmission as well.} The considered system model is illustrated in Fig. \ref{fig.sim}. The
beamforming vectors are designed to control the interference between
the cells so as to maximize a performance target \cite{CompMagcomm}.
Let us denote the set of BSs by ${\cal B}=\{1,\ldots,B\}$ and the
set of users in cell $b$ by ${\cal U}_{b}=\{1,\ldots,U\}$. User
$u$ in cell $b$ is denoted compactly as $b_{u}$. Let $s_{b_{u}}$
be an independent data symbol for user \textbf{$b_{u}$ }which is
assumed to have unit energy, i.e., $\mathbb{E}\{|s_{b_{u}}|^{2}\}=1$.
Linear transmit precoding is adopted such that the signal transmitted
to user $b_{u}$ (from BS $b$) is a multiplication of $s_{b_{u}}$
and transmit beamforming vector ${\bf v}_{b_{u}}\in\mathbb{C}^{M\times1}$.
Let ${\bf h}_{k,b_{u}}\in\mathbb{C}^{1\times M}$ denote the flat-fading
channel (row) vector between BS $k$ and user $b_{u}$. The received
signal at user $b_{u}$ can be written as
\begin{equation}
\begin{alignedat}{1}y_{b_{u}}= & \ {\bf h}_{b,b_{u}}{\bf v}_{b_{u}}s_{b_{u}}+\sum_{i\in{\cal U}_{b}\backslash\{b_{u}\}}{\bf h}_{b,b_{u}}{\bf v}_{b_{i}}s_{b_{i}}+\sum_{k\in{\cal B}\backslash\{b\}}\sum_{i\in{\cal U}_{k}}{\bf h}_{k,b_{u}}{\bf v}_{k_{i}}s_{k_{i}}+z_{b_{u}},\end{alignedat}
\label{eq:received signal}
\end{equation}
where $z_{b_{u}}$ is the additive white Gaussian noise with distribution
$z_{b_{u}}\sim{\cal CN}(0,\sigma_{b_{u}}^{2})$, $\sigma_{b_{u}}^{2}=WN_{0}$
is the noise power when using the transmission bandwidth $W$ and
the noise power spectral density is $N_{0}$. In \eqref{eq:received signal},
the second and third terms represent the intra-cell and inter-cell
interference, respectively. Let us denote by $G_{b_{u}}({\bf v})\triangleq\sum_{i\in{\cal U}_{b}\backslash\{b_{u}\}}|{\bf h}_{b,b_{u}}{\bf v}_{b_{i}}|^{2}+\sum_{k\in{\cal B}\backslash\{b\}}\sum_{i\in{\cal U}_{k}}|{\bf h}_{k,b_{u}}{\bf v}_{k_{i}}|^{2}$
the power of interference at user $b_{u}$. As is common in the system
optimization, we use the information theoretic rate expressions of
the Gaussian channels. Those assume the use of Gaussian codebooks.
Therefore, the multiuser interference terms can be modeled as additive
colored Gaussian noise and, the signal-to-interference-plus-noise
ratio (SINR) at user $b_{u}$ is expressed as
\begin{equation}
\Gamma_{b_{u}}({\bf v}) \triangleq \frac{|{\bf h}_{b,b_{u}}{\bf v}_{b_{u}}|^{2}}{G_{b_{u}}({\bf v})+\sigma_{b_{u}}^{2}}.\label{eq:SINR}
\end{equation}
The data rate of user $b_{u}$ is given by $r_{b_{u}}({\bf v})=W\log(1+\Gamma_{b_{u}}({\bf v})$),
and the total data rate over the network is given by
\begin{equation}
R({\bf v}) \triangleq \sum_{b\in{\cal B}}\sum_{u\in{\cal U}_{b}}r_{b_{u}}({\bf v}).
\end{equation}
%
%

\subsubsection*{Transmit Power Constraints}

Since the available power budget at the BSs is finite, the transmit
power at each BS should satisfy
\begin{equation}
\sum_{u\in{\cal U}_{b}}\|{\bf v}_{b_{u}}\|_{2}^{2}\leq P_{b},\forall b\in{\cal B},\label{eq:BSPower}
\end{equation}
where $P_{b}$ is the transmit power budget at BS $b$. In practice,
the power amplifier at each antenna chain is designed to operate over
a specific power range, i.e, the output power should not exceed a
predefined threshold. Thus, the power constraint for each antenna
can be also imposed, i.e.,
\begin{equation}
\sum_{u\in{\cal U}_{b}}|[{\bf v}_{b_{u}}]_{m}|^{2}\leq P_{b}^{m},\forall b\in{\cal B},\ m=1,2,...,M,\label{eq:AntennaPower}
\end{equation}
where $[{\bf x}]_{m}$ denotes the $m$th element of vector ${\bf x}$,
and $P_{b}^{m}$ is the maximum transmit power at the $m$th antenna
of BS $b$. Several other power constraints could be applied, but
we focus on these most common ones.

\subsection{Power Consumption Model}

 \begin{figure}
 \includegraphics[width=0.8\columnwidth]{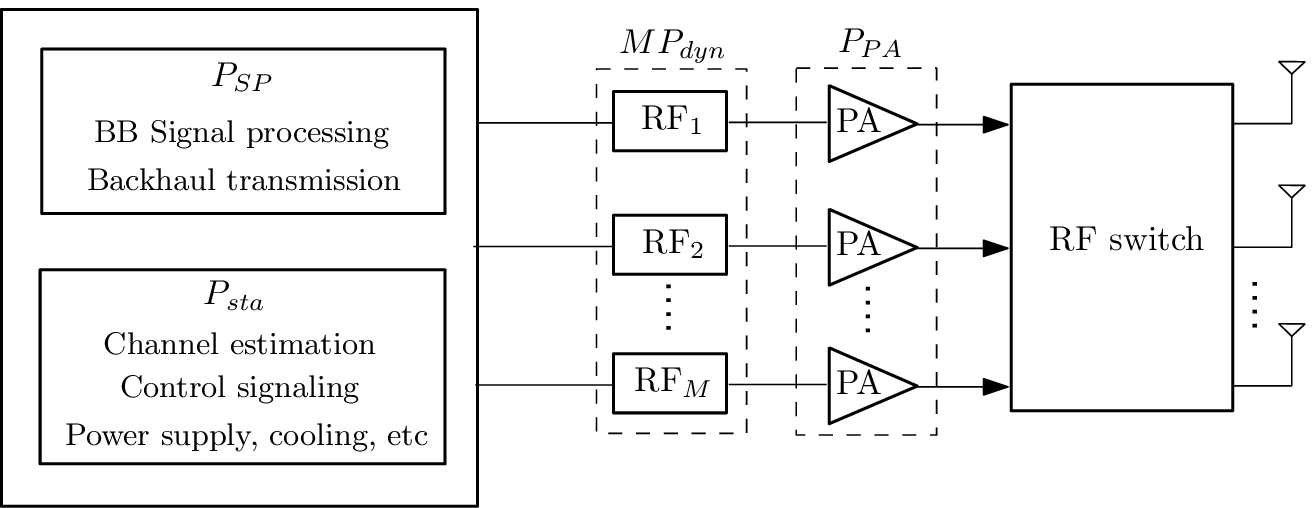}\caption{Power consumption model for a BS.}
 \label{fig.pow}
 \end{figure}
The consumed power can be classified into three main categories: circuit
operation power in network elements, signal processing power, and
power dissipated on power amplifiers (PAs). Some of the power components
are static (STA), while others are dynamic (DYN) or rate-dependent
(RD). The power consumption model is sketched in Fig. \ref{fig.pow}.

\subsubsection*{Circuit Power}

A significant amount of power is used to operate the electronic circuits
of network elements such as the base stations, site-cooling, control
signaling, backhaul infrastructure, and some parts of the baseband
processors. In addition, a radio frequency (RF) chain including, e.g.,
converters, filters and mixers requires some operating power. In general,
we can express the amount of power consumption for operating transceiver
circuits in cell $b$ as \cite{Arnold-Richter:10:PowerConsModeling,Auer2011,zappone2015energy}
\begin{equation}
P_{\text{cir},b} \triangleq P_{\text{sta}}+MP_{\text{dyn}}+UP_{\text{Us}},\label{eq:P:circuit}
\end{equation}
where $P_{\text{sta}}$ and $P_{\text{dyn}}$ represent for static
and dynamic power consumption at BS $b$, respectively, and $P_{\text{Us}}$
accounts for power running a user device.

\subsubsection*{Signal Processing Power}

The data needs to be encoded and modulated at the
transmitter as well as demodulated and decoded at the receiver. Conventionally,
the amount of power for these functionalities is assumed to be fixed
\cite{zappone2015energy,DerrickKwanNg:2012:JWCOM:EE_OFDM,She:2014:WSEE,HHJY:13:JCOM,HHJYL:13:JCOML}.
However, generally, a higher data rate requires a larger codebook,
and the larger number of bits incurs higher power for encoding and
decoding on baseband circuit boards. Moreover, the backhaul is used
to transmit data between the core network and the BSs, and the power
consumed for the backhaul also increases with the data rate \cite{Isheden:2010:globecom,Xiong:2012:EEOFDMA,Wang:2013:EERate-dependentPower}.
From this perspective, signal processing power consumption is rate-dependent,
and is assumed to be a linear function of the transmission rate \cite{Isheden:2010:globecom}.
Let us denote by $P_{\text{SP},b}(r_{b}({\bf v}))$, where $r_{b}({\bf v}) \triangleq \sum_{u\in{\cal {U}}_{b}}r_{b_{u}}({\bf v})$,
the signal processing power for BS $b$. Then we can write
\begin{equation}
P_{\text{SP},b}(r_{b}({\bf v}))\triangleq\begin{cases}
\textrm{constant} & \textrm{for\ fixed\ power\ model}\\
{\displaystyle p_{\text{SP}}\sum_{u\in{\cal {U}}_{b}}r_{b_{u}}({\bf v})} & \textrm{for\ rate-dependent\ power\ model}
\end{cases},\label{eq:P:RD}
\end{equation}
where $p_{\text{SP}}$ is a constant coefficient with unit W/(Gbits/s).

\subsubsection*{Power Dissipated on PAs}

The amount of power consumed by the PAs strongly
depends on the power amplifier's efficiency. Conventionally, the efficiency
of a PA is assumed to be a constant over operating range \cite{zappone2015energy,DerrickKwanNg:2012:JWCOM:EE_OFDM,She:2014:WSEE,HHJY:13:JCOM,HHJYL:13:JCOML}.
This assumption leads to the model
\begin{equation}
P_{\text{PA},b}({\bf v}) \triangleq \frac{1}{\epsilon}\sum_{u\in{\cal {U}}_{b}}\|{\bf v}_{b_{u}}\|_{2}^{2},\label{eq:P:PA:linear}
\end{equation}
where $P_{\text{PA},b}$ denotes the PAs' dissipated power at BS $b$,
and $\epsilon\in(0,1)$ is a constant standing for the PA efficiency.
However, in practice, PA efficiency is highly dependent on the output
power region and the employed PA type. To account this, the non-linear
power consumption models of PAs have been introduced \cite{Mikami2007,Bjoernemo2009,AmplifierMIMO-Persson,Tervo:nonlinearPA}
in which the PA efficiency of RF chain $m$ at BS $b$ is expressed
as
\begin{equation}
\text{\ensuremath{\epsilon}}_{b,m}(\{{\bf v}_{b_{u}}\}_{u}) \triangleq \tilde{\epsilon}\sqrt{\sum_{u\in{\cal {U}}_{b}}|[{\bf v}_{b_{u}}]_{m}|^{2}},\label{eq:nonlinear-PA-Eff}
\end{equation}
where $\tilde{\epsilon}=\epsilon_{\max}/\sqrt{P_{b}^{m}}$, and
$\epsilon_{\max}\in(0,1)$ is the maximum PA's efficiency. We note
that $P_{b}^{m}$ and $\epsilon_{\max}$ depend on the employed PA
techniques. For notational simplicity, we assume that $\tilde{\epsilon}$
is the same for all $b$, $m$. From \eqref{eq:nonlinear-PA-Eff},
the total power consumption on the PAs at BS $b$ can be written as
\begin{equation}
P_{\text{PA},b}({\bf v}) \triangleq \sum_{m=1}^{M}\frac{\sum_{u\in{\cal {U}}_{b}}|[{\bf v}_{b_{u}}]_{m}|^{2}}{\text{\ensuremath{\epsilon}}_{b,m}(\{{\bf v}_{b_{u}}\}_{u})}=\frac{1}{\tilde{\epsilon}}\sum_{m=1}^{M}\sqrt{\sum_{u\in{\cal {U}}_{b}}|[{\bf v}_{b_{u}}]_{m}|^{2}}.\label{eq:P:PA:nonlinear}
\end{equation}

\subsubsection*{General Power Consumption Models}

Based on the above discussion, the total power consumption model in
cell $b$ can be collectively written as
\begin{equation}
P_{\text{BS},b}({\bf v}) \triangleq P_{\text{cir},b}+P_{\text{SP},b}(r_{b}({\bf v}))+P_{\text{PA},b}({\bf v}).\label{eq:gen:PowerCons:BS}
\end{equation}
Hence, the total network power consumption is
\begin{equation}
P_{\text{total}}({\bf v}) \triangleq \sum_{b\in{\cal B}}\left(P_{\text{cir},b}+P_{\text{SP},b}(r_{b}({\bf v}))+P_{\text{PA},b}({\bf v})\right).\label{eq:gen:PowerCons:total}
\end{equation}
On the other hand, the power for the data transmission to a user is
a favorable measure in some user-centric applications. Let
\[
P_{\text{SP},b}(r_{b_{u}}({\bf v}))\triangleq\begin{cases}
P_{\text{SP},b}(r_{b}({\bf v}))/U & \textrm{for\ fixed\ power\ model}\\
{\displaystyle p_{\text{SP}}r_{b_{u}}({\bf v})} & \textrm{for\ rate-dependent\ power\ model}
\end{cases}
\]
denote the signal processing power corresponding to user $b_{u}$.
Then the amount of consumed power corresponding to the data transmission
to user $b_{u}$ can be written as \cite{HHJYL:13:JCOML}
\begin{equation}
P_{\text{Us},b_{u}}({\bf v}) \triangleq \sum_{m=1}^{M} \frac{|[{\bf v}_{b_{u}}]_{m}|^{2}}{\tilde{\epsilon}\sqrt{\sum_{u\in{\cal {U}}_{b}}|[{\bf v}_{b_{u}}]_{m}|^{2}}}+P_{\text{SP},b}(r_{b_{u}}({\bf v}))+\frac{P_{\text{sta}}+MP_{\text{dyn}}}{U}+P_{\text{Us}}\label{eq:gen:PowerConsUser}
\end{equation}
in which all users in a cell are assumed to be evenly responsible
for the operating power of their serving BS.

\subsection{Energy-Efficiency Metrics\label{sec:Network-centric-Energy-Efficient}}

\begin{figure}
 \centering{}\includegraphics[width=0.8\columnwidth]{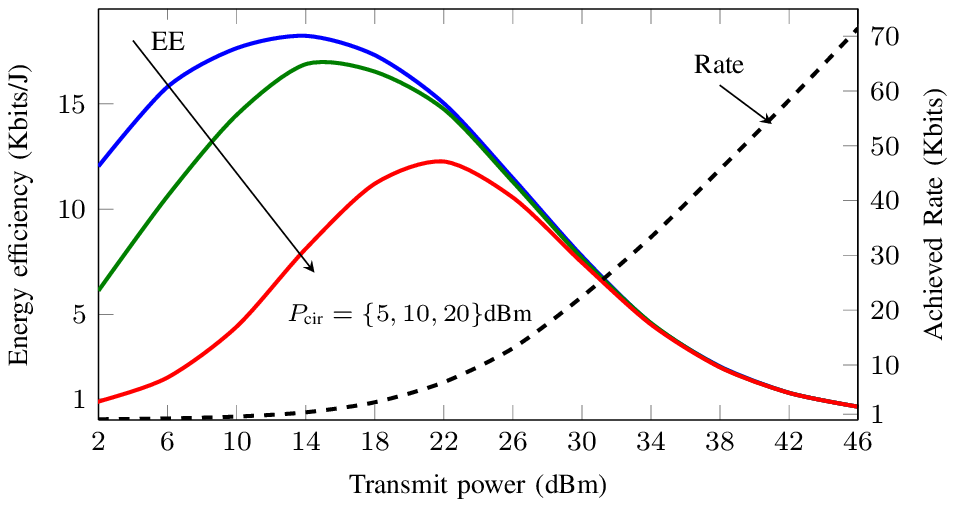}\caption{Energy efficiency (solid curves) and the achieved user rate (dashed curve) versus the transmit power for single-cell single-user MISO downlink. The simulation parameters  are given in Table I.}
 \label{fig.ee-rate}
 \end{figure}

The EE measures the number of bits reliably transmitted by a unit
energy. In other words, it can be defined as the ratio of the achievable
data rate to the total power consumption. The ratio quantifies
the trade-off between the network throughput and the power consumption.
This is illustrated via a simple single-cell single-user MISO downlink
example in Fig. \ref{fig.ee-rate}. The energy efficiency and the achieved
rate are plotted versus the transmit power. We observe that, for all
cases of the operating circuit power, when the transmit power increases,
the EE first increases, reaching a maximum, and then decreases. In
other words, when the circuit power plays a non-negligible role and
the rate is penalized by the overall power consumption, the optimum
performance is not achieved by using all available power budget. This
observation gives rise to the systematic development of the optimization
algorithms as detailed below, where four different widely considered
EE metrics are introduced and discussed.

\subsubsection*{Network Energy Efficiency}

The NEE metric quantifies the EE performance of the entire network
\cite{zappone2015energy,Oskari:2017:TSP}. It is defined as
\begin{equation}
\mathrm{NEE}({\bf v})\triangleq\frac{R({\bf v})}{P_{\text{total}}({\bf v})}\label{eq:NEE:define}
\end{equation}
We remark that in scenarios where cellular BSs with different features
and specifications co-exist, e.g., heterogeneous network, NEE might
lack relevance, since neither EE requirement for each cell/user nor
the fairness among all parties of the network can be guaranteed.

\subsubsection*{Sum Weighted Energy Efficiency}

The SWEE metric can satisfy the specific demand on EE of each network
node. For the considered system model, SWEE can be expressed as
\begin{equation}
\mathrm{SWEE}({\bf v})\triangleq\begin{cases}
{\displaystyle \sum_{b\in{\cal {B}}}\omega_{b}\frac{\sum_{u\in{\cal {U}}_{b}}r_{b_{u}}({\bf v})}{P_{\text{BS,}b}({\bf v})}} & \textrm{for\ the\ BS-centric\ network}\\
{\displaystyle \sum_{b\in{\cal {B}}}\sum_{u\in{\cal {U}}_{b}}\omega_{b_{u}}\frac{r_{b_{u}}({\bf v})}{P_{\text{Us},b_{u}}({\bf v})}} & \textrm{for\ the\ user-centric\ network}
\end{cases}\label{eq:SWEE:define}
\end{equation}
where $\omega_{b}\in(0,1]$ and $\omega_{b_{u}}\in(0,1]$ are parameters
representing the priority for cell $b$ and user $b_{u}$, respectively.

\subsubsection*{Weighted Product EE}

The WPEE metric also takes into account the individual demand on EE
of each node which is defined as \cite{zappone2015energy,Venturino:2015:TWC}
\begin{equation}
\mathrm{WPEE}({\bf v})\triangleq\begin{cases}
{\displaystyle \prod_{b\in{\cal {B}}}\left(\frac{\sum_{u\in{\cal {U}}_{b}}r_{b_{u}}({\bf v})}{P_{\text{BS,}b}({\bf v})}\right)^{\omega_{b}}} & \textrm{for\ the\ BS-centric\ network}\\
{\displaystyle \prod_{b\in{\cal {B}}}\prod_{u\in{\cal U}_{b}}\left(\frac{r_{b_{u}}({\bf v})}{P_{\text{Us},b_{u}}({\bf v})}\right)^{\omega_{b_{u}}}} & \textrm{for\ the\ user-centric\ network}
\end{cases}
\end{equation}
It is worth noting that although the WPEE metric does not give the
same EE-unit (bits/J) as such, it has been used in the literature
to achieve fairness in EE. Specifically, it is not difficult to see
that none of the BSs experiences EE close to zero when
WPEE is considered.

\subsubsection*{Max-min Fairness Energy Efficiency}

The max-min fairness EE metric provides the best fairness for the
considered nodes compared to the others. This metric is preferable
to the scenarios where EE is critical for each cell, e.g., in cellular
networks where BSs are not connected to fixed electricity grid. The
definition of the metric is given as \cite{Giang:15:JCOML}
\begin{align}
\mathrm{minEE}({\bf v})\triangleq\begin{cases}
{\displaystyle \min_{b\in{\cal B}}\frac{\sum_{u\in{\cal {U}}_{b}}r_{b_{u}}({\bf v})}{P_{\text{BS,}b}({\bf v})}} & \textrm{for\ the\ BS-centric\ network}\\
{\displaystyle \min_{\substack{b\in{\cal B}\\
u\in{\cal {U}}_{b}
}
}\frac{r_{b_{u}}({\bf v})}{P_{\text{Us},b_{u}}({\bf v})}} & \textrm{for\ the\ user-centric\ network}
\end{cases}\label{eq:max-minEE}
\end{align}

\subsection{Energy Efficiency Optimization Problems\label{sec:Network-centric-Energy-Efficient-Problem}}

From the above discussions, the problems of beamforming design for
EE maximization can be generally written as
\begin{equation}\label{eq:EEgeneral}
\underset{{\bf v}}{\text{maximize}}\ f_{\mathrm{EE}}({\bf v})\quad\text{subject to}\ \bigl\{\eqref{eq:BSPower},\eqref{eq:AntennaPower}\bigr\}
\end{equation} where the objective function $f_{\mathrm{EE}}({\bf v})$
represents one of $\mathrm{NEE}({\bf v})$, $\mathrm{SWEE}({\bf v})$,
$\mathrm{WPEE}({\bf v})$, and $\mathrm{minEE}({\bf v})$.

In general, \eqref{eq:EEgeneral} is a nontractable fractional program.\footnote{By nontractable we mean that it cannot be reformulated as an equivalent
convex program or such a transformation is not known in the literature.} In the next section, we briefly review conventional approaches, which
suboptimally solve the EEmax problems, and then provide the recently
proposed SCA framework which improves efficiently solution quality.

\section{Centralized Methods for Energy-efficient Transmissions}

\label{sec:centraldistributedsols}

\subsection{Conventional Fractional Programming Approaches\label{subsec:Conventional-Approaches}}

Most of existing solutions for the EEmax problems are based on
conventional fractional programming methods, i.e., parameterized approaches
\cite{zappone2015energy,DerrickKwanNg:2012:JWCOM:EE_OFDM,She:2014:WSEE}
or the parameter-free approach based on the Charnes-Cooper transformation.
We briefly sketch the idea of these approaches for solving fractional
programs below.

In general, a fractional program is expressed as
\begin{align}
\underset{{{\bf x}}\in{\cal S}}{\text{maximize}}\  & \sum_{i=1}^{L}\frac{f_{i}({\bf {x})}}{g_{i}({\bf {x})}}\label{FP:general}
\end{align}
where $L\geq1$, ${\cal S}$, $f_{i}({\bf {x})}$ and $g_{i}({\bf {x})}$
are a convex set, concave and convex functions respect to variable
vector ${\mathbf{x}\in\mathbb{C}^{N}}$, respectively.

\subsubsection*{Single-ratio Fractional Programs}

When $L=1$, the problem can be transformed into a parameterized form.
That is, one can consider the following problem with parameter $\omega$:
$H(\omega)={\displaystyle \max_{{\bf {x}\in{\cal S}}}}\{f_{1}(\mathbf{x})-\omega g_{1}(\mathbf{x})\}$.
Due to the fact that $H(\omega)$ is continuous and strictly monotonically
decreasing \cite{dinkelbach1967nonlinear}, $H(\omega)=0$ has a unique
solution $\omega^{\ast}$. The optimal solution to the original fractional
program is $\mathbf{x}^{\ast}={\displaystyle \arg\max_{{\bf {x}\in{\cal S}}}}\{f_{1}(\mathbf{x})-\omega^{\ast}g_{1}(\mathbf{x})\}$.
Thereby, the problem can be solved by finding $\omega$ such that
$H(\omega)=0$. A parametric approach exploits the Newton method to
find root of $H(\omega)$ (often called as the Dinkelbach method or
the Newton-Rhapson method). The method first initializes $\omega^{(0)}=\frac{f_{1}(\mathbf{x}^{(0)})}{g_{1}(\mathbf{x}^{(0)})}$.
Subsequently, the problem $H(\omega^{(0)})$ is solved, the solution
of which is then used to update $\omega^{(n)}=\frac{f_{1}(\mathbf{x}^{(n-1)})}{g_{1}(\mathbf{x}^{(n-1)})}$,
and this procedure is repeated until convergence. Besides the well-known
Dinkelbach method, the problem can be also solved as a single convex
program using the Charnes-Cooper transformation \cite{schaible1976fractional}.

\subsubsection*{Multi-ratio Fractional Programs}

When $L>1$, \eqref{FP:general} is a sum-of-ratios fractional program.
A conventional heuristic strategy for solving this type of problems with concave-convex
ratios is to transform it to a parameterized form with some fixed
parameters, and then search the optimal parameters by solving a series
of convex subproblems \cite{She:2014:WSEE,Wu-16,GYu:2015:TWCOM}. Specifically, the
solutions for \eqref{FP:general} can be obtained by solving as a
series of subproblems $H(\boldsymbol{\alpha},\boldsymbol{\beta})={\displaystyle \max_{{\bf {x}\in{\cal S}}}}\sum_{i=1}^{L}\alpha_{i}(f_{i}(\mathbf{x})-\beta_{i}g_{i}(\mathbf{x}))$,
where $\{\alpha_{i}\}_{i},\{\beta_{i}\}_{i}$ are parameters. Similarly
to the single-ratio case, $\{\alpha_{i}\}_{i},\{\beta_{i}\}_{i}$
are first fixed and the subproblem is solved for given parameters.
Then, $\{\alpha_{i}\}_{i},\{\beta_{i}\}_{i}$ are updated according
to a damped Newton method. 

Nevertheless, the advantages of the parametric approaches are hardly
recognized when they are applied to wireless communications problems
because $f_{i}(\mathbf{x})$ and $g_{i}(\mathbf{x})$ are often non-convex.
Implicitly, the parametric subproblem is nonconvex and its optimal
solutions are difficult to find. To cope with this, the SCA or alternating
optimization method based on iterative weighted minimum mean square
error (WMMSE) approach is often combined with the parametric method
leading to multi-level iterative algorithms. Thus, these algorithms
need a very high number of iterations to converge. Moreover, likely
local optimality for each parametric problem is achieved which means
that parametric approaches may not always guarantee the convergence.

To avoid the multi-level iterative procedure, we present below the framework developed recently based on the SCA
method. The algorithms derived from the approach are provably and
fast convergent, thus, they overcome the issues raised by the earlier
solutions.

\subsection{SCA Principle \label{subsec:SCA-Principle}}

We first briefly review the SCA principles before presenting their
applications to the EEmax problems. The central idea of the SCA method
is to iteratively approximate the nonconvex constraints of an optimization
problem by proper convex ones \cite{MarksWright:78:AGenInnerApprox}.
In particular, let us consider a general optimization program given
by
\begin{equation}
\underset{{\bf x}\in{\cal S}}{\text{minimize}}\enskip f({\bf x})\quad\text{subject to}\quad\{g_{i}({\bf x})\leq0,\ i=1,...,L\}\label{eq:SCA:gen}
\end{equation}
where $f({\bf x})$ is convex and $\{g_{i}({\bf x})\}_{i}$ are nonconvex
functions in a convex set ${\cal S}$ w.r.t variable vector ${\bf x}$.
At iteration $n$, given a feasible point ${\bf x}^{(n)}$, function
$g_{i}({\bf x})$ is approximated by its convex approximation function
$\hat{g}_{i}({\bf x},{\bf x}^{(n)})$ for all $i$ such that
\begin{description}
\item [{(a)}] $g_{i}({\bf x})\leq\hat{g}_{i}({\bf x},{\bf x}^{(n)})$
\item [{(b)}] $g_{i}({\bf x}^{(n)})=\hat{g}_{i}({\bf x}^{(n)},{\bf x}^{(n)})$
\item [{(c)}] $\nabla_{{\bf x}}g_{i}({\bf x}^{(n)})=\nabla_{{\bf x}}\hat{g}_{i}({\bf x}^{(n)},{\bf x}^{(n)})$
\end{description}
for all ${\bf x}\in\tilde{{\cal S}}\triangleq\{{\bf x}\in{\cal S}|g_{i}({\bf x})\leq0,\ i=1,...,L\}$.
Properties (a) and (b) are to guarantee the monotonic (objective)
convergence behavior for the SCA algorithm; properties (b) and (c)
guarantee that the Karush–Kuhn–Tucker (KKT) optimality conditions are satisfied by convergent
points\cite{MarksWright:78:AGenInnerApprox}. By the replacement,
we arrive at the following convex subproblem
\begin{equation}
\underset{{\bf x}\in{\cal S}}{\text{minimize}}\enskip f({\bf x})\quad\text{subject to}\quad\{\hat{g}_{i}({\bf x},{\bf x}^{(n)})\leq0,\ i=1,...,L\}.\label{eq:SCA:approx}
\end{equation}
The optimal solution ${\bf x}^{\ast}$ of \eqref{eq:SCA:approx} belongs
to the set $\tilde{{\cal S}}$ due to (a) and (b). Thus, ${\bf x}^{\ast}$
is used as the feasible point for the next iteration, i.e. ${\bf x}^{(n+1)}={\bf x}^{\ast}$.
The process is iteratively carried out until convergence is established.
The SCA procedure solving \eqref{eq:SCA:gen} is outlined in Algorithm
\ref{Alg.SCA}. We note that $f({\bf x}^{\ast})\leq f({\bf x}^{(n)})$
for all $n$, i.e. sequence $\{f({\bf x}^{(n)})\}_{n}$ decreases
monotonically. Thus, $\{f({\bf x}^{(n)})\}_{n}$ converges if it is
bounded below by a finite value in the set $\tilde{{\cal S}}$. The
following remark shows a well-known method for arriving $\hat{g}_{i}({\bf x},{\bf x}^{(n)})$,
which is widely used in this paper.
\begin{algorithm}[tb]
\caption{SCA Procedure Solving \eqref{eq:SCA:gen}}
\label{Alg.SCA}

\begin{algorithmic}

\STATE \textbf{Initialization:} Set $n:=0$,\textbf{ }choose an initial
feasible point ${\bf x}^{(n)}$.

\REPEAT

\STATE{Solve \eqref{eq:SCA:approx} and obtain optimal value ${\bf x}^{\ast}$}

\STATE{Update ${\bf x}^{(n+1)}:={\bf x}^{\ast}$}

\STATE{Update $n:=n+1$}

\UNTIL{Convergence}

\STATE{\textbf{Output}: ${\bf x}^{(n)}$}

\end{algorithmic}
\end{algorithm}

\begin{rem}
Let $g({\bf x})$ be a concave function w.r.t ${\bf x}$ , then its
convex upper bound satisfying (a)\textendash (c) can be achieved by
the mean of the first order Taylor approximation as
\begin{align}
g({\bf x}) & \leq\underset{\triangleq\hat{g}({\bf x};{\bf x}^{(n)})}{\underbrace{g({\bf x}^{(n)})+\left\langle \nabla_{{\bf x}}g({\bf x}^{(n)}),{\bf x}-{\bf x}^{(n)}\right\rangle }}.\label{eq:first-order-app}
\end{align}
\end{rem}
\begin{example}
\label{examp:appr} Consider the quadratic-over-linear function $g(x,y)=\frac{-x^{2}}{y}$,
$y>0$, which is concave w.r.t the involved variables. From \eqref{eq:first-order-app},
a convex upper bound of $g(x,y)$ at $(x^{(n)},y^{(n)})$, $y^{(n)}>0$,
is written as
\begin{align*}
\hat{g}(x,y;x^{(n)},y^{(n)}) & =-\left(\frac{\left(x^{(n)}\right)^{2}}{y^{(n)}}+\frac{2x^{(n)}}{y^{(n)}}(x-x^{(n)})-\frac{(x^{(n)})^{2}}{(y^{(n)})^{2}}(y-y^{(n)})\right)\\
 & =-\frac{2x^{(n)}}{y^{(n)}}x+\frac{(x^{(n)})^{2}}{(y^{(n)})^{2}}y.
\end{align*}
It can be easily justified that $\hat{g}(x,y;x^{(n)},y^{(n)})$ satisfies
properties (a)\textendash (c) for all $(x,y>0)$.
\end{example}

\subsection{SCA based Solutions for EEmax Problems\label{subsec:SCA-based-Solutions}}

In this subsection, we present how to adopt the procedure discussed
in Subsection \ref{subsec:SCA-Principle} to the EEmax problems posted
in Subsection \ref{sec:Network-centric-Energy-Efficient-Problem}.
It is worth mentioning that directly applying the SCA method to these
problems seems challenging, because deriving convex approximations for
nonconvex parts in the problems that satisfy conditions (a)\textendash (c)
is very difficult. Thus, the necessary step is to transform the EEmax
problems into more tractable representations, which preserve the optimality
of the original one as well as are amenable to the SCA method.

\subsubsection{Network EEmax Problem}

We first provide the SCA solutions for the problem with network EE
metric which contains single-ratio fractional objective. Replacing
$f_{\mathrm{EE}}({\bf v})$ in \eqref{eq:EEgeneral} by $\mathrm{NEE}({\bf v})$,
we get the following problem
\begin{equation}
\underset{{\bf v}}{\text{maximize}}\enskip\frac{R({\bf v})}{P_{\text{total}}({\bf v})}\quad\text{subject to}\enskip\bigl\{\eqref{eq:BSPower},\eqref{eq:AntennaPower}\bigr\}.\label{Prob:NEE}
\end{equation}
For translating \eqref{Prob:NEE} to a more tractable form, we exploit
the epigraph transformation \cite{boyd2004convex}. Let us introduce
new slack variables $\eta$, $z$, $t$ and $\{g_{b_{u}}\}_{b_{u}}$
and rewrite \eqref{Prob:NEE} as \begin{subequations}\label{Prob:NEE:epi}
\begin{align}
\underset{{\bf v},\eta,z,t,\{g_{b_{u}}\}}{\text{maximize}} & \quad\eta\\
\text{subject to} & \quad\eta\leq\frac{z^{2}}{t}\label{eq:NEE:EE:epi}\\
 & \quad t\geq P_{\text{total}}({\bf v})\label{eq:NEE:PowerCons:epi}\\
 & \quad z^{2}\leq\sum_{b\in{\cal B},u\in{\cal U}_{b}}\log(1+g_{b_{u}})\label{eq:NEE:rate:epi}\\
 & \quad\frac{|{\bf h}_{b,b_{u}}{\bf v}_{b_{u}}|^{2}}{g_{b_{u}}}\geq G_{b_{u}}({\bf v})+\sigma_{b_{u}}^{2},\ \forall b\in{\cal B},u\in{\cal U}_{b}\label{eq:NEE:SINR:epi}\\
 & \quad\eqref{eq:BSPower},\eqref{eq:AntennaPower}.
\end{align}
\end{subequations}The relationship between \eqref{Prob:NEE} and
\eqref{Prob:NEE:epi} is stated in the following lemma.
\begin{lem}
\label{lem:equivalence}Problems \eqref{Prob:NEE} and \eqref{Prob:NEE:epi}
are equivalent at optimality.
\end{lem}
The proof of the lemma is given in Appendix.
Let us now apply the SCA method to solve \eqref{Prob:NEE:epi}. First,
we observe that constraints \eqref{eq:NEE:EE:epi} and \eqref{eq:NEE:SINR:epi}
are nonconvex while the others are convex. Second, the nonconvex parts
in \eqref{eq:NEE:EE:epi} and \eqref{eq:NEE:SINR:epi} are in the
form of quadratic-over-affine function mentioned in Example \ref{examp:appr}.
Therefore, the valid convex approximations for \eqref{eq:NEE:EE:epi}
and \eqref{eq:NEE:SINR:epi} are given as
\begin{align}
\frac{2z^{(n)}}{t^{(n)}}z-\frac{(z^{(n)})^{2}}{(t^{(n)})^{2}}t & \geq\eta\label{eq:approx:1}\\
\frac{2\Re(\tilde{\mathbf{h}}_{b,b_{u}}^{(n)}\mathbf{v}_{b_{u}})}{g_{b_{u}}^{(n)}}-\frac{|\mathbf{h}_{b,b_{u}}\mathbf{v}_{b_{u}}^{(n)}|{}^{2}g_{b_{u}}}{(g_{b_{u}}^{(n)})^{2}} & \geq G_{b_{u}}({\bf v})+\sigma_{b_{u}}^{2}\label{eq:approx:2}
\end{align}
respectively, where $\tilde{\mathbf{h}}_{b,b_{u}}^{(n)}\triangleq(\mathbf{v}_{b_{u}}^{(n)}){}^{H}\mathbf{h}_{b,b_{u}}^{H}\mathbf{h}_{b,b_{u}}$
and $({\bf v}^{(n)},z^{(n)},t^{(n)},\{g_{b_{u}}^{(n)}\})$ is some
feasible point of \eqref{Prob:NEE:epi}. As a result, we arrive at
the approximate convex program at iteration $n$ as
\begin{align}
\underset{{\bf v},\eta,z,t,\{g_{b_{u}}\}}{\text{maximize}} & \enskip\eta\quad\text{subject to}\ \{\eqref{eq:BSPower},\eqref{eq:AntennaPower},\eqref{eq:NEE:PowerCons:epi},\eqref{eq:NEE:rate:epi},\eqref{eq:approx:1},\eqref{eq:approx:2}\}.\label{Prob:NEE:approx}
\end{align}
For the rate-dependent signal processing
model, due to the following relation $\arg\max_{{\bf v}}\frac{R({\bf v})}{P_{\text{total}}({\bf v})}=\arg\min_{{\bf v}}\frac{P_{\text{total}}({\bf v})}{R({\bf v})}=\arg\min_{{\bf v}}\frac{\sum_{b\in{\cal B}}\left(P_{\text{cir},b}+P_{\text{PA},b}({\bf v})\right)}{R({\bf v})}+p_{\text{SP}}=\arg\max_{{\bf v}}\frac{R({\bf v})}{\sum_{b\in{\cal B}}\left(P_{\text{cir},b}+P_{\text{PA},b}({\bf v})\right)}$, we can ignore the term of rate-dependent power in the optimization process without loss of optimality.
Consequently, the denominator of the objective becomes a convex function
w.r.t. ${\bf v}$, and thus the solutions can be obtained
following the above discussion.

\subsubsection{Sum Weighted EEmax Problem}

We focus on the problem of SWEE maximization from the perspective
of the BSs. The SWEE maximization problem from the user perspective
is treated similarly. Replacing $f_{\mathrm{EE}}({\bf v})$ in \eqref{eq:EEgeneral}
by $\mathrm{SWEE}({\bf v})$, we arrive at the problem
\begin{equation}
\underset{{\bf v}}{\text{maximize}}\enskip\sum_{b=1}^{B}\omega_{b}\frac{\sum_{u\in{\cal {U}}_{b}}r_{b_{u}}({\bf v})}{P_{\text{BS,}b}({\bf v})}\quad\text{subject to}\enskip\bigl\{\eqref{eq:BSPower},\eqref{eq:AntennaPower}\bigr\}.\label{Prob:SWEE}
\end{equation}
As the first step, we introduce new variables $\{\eta_{b}\}_{b},\{z_{b}\}_{b},\{t_{b}\}_{b},\{g_{b_{u}}\}_{b_{u}}$
and write \eqref{Prob:SWEE} in equivalent form as \begin{subequations}\label{Prob:SWEE:epi}
\begin{align}
\underset{{\bf v},\{\eta_{b}\},\{z_{b}\},\{t_{b}\},\{g_{b_{u}}\}}{\text{maximize}} & \quad\sum_{b\in{\cal B}}\omega_{b}\eta_{b}\label{eq:obj:SWEE:epi}\\
\text{subject to} & \quad\eta_{b}\leq\frac{z_{b}^{2}}{t_{b}},\ \forall b\in{\cal B}\label{eq:SWEE:EE:epi}\\
 & \quad t_{b}\geq P_{\text{BS},b}({\bf v}),\ \forall b\in{\cal B}\label{eq:SWEE:PowerCons:epi}\\
 & \quad z_{b}^{2}\leq\sum_{u\in{\cal U}_{b}}\log(1+g_{b_{u}}),\ \forall b\in{\cal B}\label{eq:SWEE:rate:epi}\\
 & \quad\frac{|{\bf h}_{b,b_{u}}{\bf v}_{b_{u}}|^{2}}{g_{b_{u}}}\geq G_{b_{u}}({\bf v})+\sigma_{b_{u}}^{2},\ \forall b\in{\cal B},u\in{\cal U}_{b}\label{eq:SWEE:SINR:epi}\\
 & \quad\eqref{eq:BSPower},\eqref{eq:AntennaPower}.
\end{align}
\end{subequations}The equivalence between \eqref{Prob:SWEE} and
\eqref{Prob:SWEE:epi} can be easily justified following the procedure
in the proof for Lemma \ref{lem:equivalence}. The nonconvex parts
of problem \eqref{Prob:SWEE:epi} lie in \eqref{eq:SWEE:EE:epi} and
\eqref{eq:SWEE:SINR:epi} which can be approximated in convex forms as
\begin{align}
\frac{2z_{b}^{(n)}}{t_{b}^{(n)}}z_{b}-\frac{(z_{b}^{(n)})^{2}}{(t_{b}^{(n)})^{2}}t_{b} & \geq\eta_{b},\ \forall b\in{\cal B}\label{eq:approx:1-1}\\
\frac{2\Re(\tilde{\mathbf{h}}_{b,b_{u}}^{(n)}\mathbf{v}_{b_{u}})}{q_{b_{u}}^{(n)}}-\frac{|\mathbf{h}_{b,b_{u}}\mathbf{v}_{b_{u}}^{(n)}|{}^{2}q_{b_{u}}}{(q_{b_{u}}^{(n)})^{2}} & \geq G_{b_{u}}({\bf v})+\sigma_{b_{u}}^{2},\ \forall b\in{\cal B},u\in{\cal U}_{b}\label{eq:approx:2-1}
\end{align}
respectively. Then, the subproblem solved in iteration $n$ is
\begin{align}
\underset{{\bf v},\{\eta_{b}\},\{z_{b}\},\{t_{b}\},\{g_{b_{u}}\}}{\text{maximize}} & \enskip\sum_{b\in{\cal B}}\omega_{b}\eta_{b}\quad\text{subject to}\ \{\eqref{eq:BSPower},\eqref{eq:AntennaPower},\eqref{eq:SWEE:PowerCons:epi},\eqref{eq:SWEE:rate:epi},\eqref{eq:approx:1-1},\eqref{eq:approx:2-1}\}.\label{Prob:SWEE:approx}
\end{align}
For the rate-dependent signal processing model, we replace constraints
in \eqref{eq:SWEE:rate:epi} and \eqref{eq:SWEE:PowerCons:epi} by
\[
\begin{array}{rl}
t_{b}+z_{b}^{2}p_{\text{SP}} & \leq\frac{z_{b}^{2}}{\eta_{b}},\ \forall b\in{\cal B}\\
t_{b} & \geq P_{\text{cir},b}+P_{\text{PA},b}({\bf v}),\ \forall b\in{\cal B}.
\end{array}
\]
The same transformation can be applied also to the following problems.

\subsubsection{Weighted Product EEmax Problem}

WPEE metric has been considered in power control problems so far \cite{zappone2015energy,Venturino:2015:TWC}.
However, to the best of our knowledge, beamforming designs for WPEE
maximization have not been yet investigated. We show below that the
proposed framework can be straightforwardly applied to the problem
with such metric. The problem of beamforming designs for WPEE maximization
reads
\begin{equation}
\underset{{\bf v}}{\text{maximize}}\enskip\prod_{b\in{\cal {B}}}\left(\frac{\sum_{u\in{\cal {U}}_{b}}r_{b_{u}}({\bf v})}{P_{\text{BS,}b}({\bf v})}\right)^{\omega_{b}}\quad\text{subject to}\enskip\bigl\{\eqref{eq:BSPower},\eqref{eq:AntennaPower}\bigr\}\label{Prob:PWEE}
\end{equation}
Also, we introduce new slack variables $\{\eta_{b}\},\{z_{b}\},\{t_{b}\},\{g_{b_{u}}\}$,
then translate \eqref{Prob:PWEE} into a more tractable form given
as\begin{align}\label{Prob:PWEE:epi}
\underset{{\bf v},\{\eta_{b}\},\{z_{b}\},\{t_{b}\},\{g_{b_{u}}\}}{\text{maximize}}\enskip\prod_{b\in{\cal {B}}}(\eta_{b})^{\omega_{b}}\quad\text{subject to}\enskip\bigl\{\eqref{eq:BSPower},\eqref{eq:AntennaPower},\eqref{eq:SWEE:EE:epi},\eqref{eq:SWEE:PowerCons:epi},\eqref{eq:SWEE:rate:epi},\eqref{eq:SWEE:SINR:epi}\bigr\}
\end{align}Again, we can justify the equivalence between \eqref{Prob:PWEE}
and \eqref{Prob:PWEE:epi} at the optimum similar to that for Lemma
\ref{lem:equivalence}. We note that the objective function of \eqref{Prob:PWEE:epi}
is generally neither concave nor convex since the exponents $\{\omega_{b}\}_{b}$
are arbitrary positive values. A simple way to overcome the issue
is to scale the exponents so that the objective function turns into
a concave monomial function which is conic quadratic representable
\cite{BenNemi:book:LectModConv}. Particularly, we can always find
$\alpha>1$ such that $\tilde{\omega}_{b}=\frac{\omega_{b}}{\alpha}$
for all $b$ and $\sum_{b}\tilde{\omega}_{b}\leq1$. Then $\prod_{b\in{\cal {B}}}(\eta_{b})^{\tilde{\omega}_{b}}$
is concave monomial. We also note that the optimal solution to \eqref{Prob:PWEE:epi}
stays the same under the scale. Now, we are ready to arrive at the
convex subproblem solved at iteration $n$ of the SCA algorithm given
as

\begin{align}
\underset{{\bf v},\{\eta_{b}\},\{z_{b}\},\{t_{b}\},\{g_{b_{u}}\}}{\text{maximize}} & \quad\prod_{b\in{\cal {B}}}(\eta_{b})^{\tilde{\omega}_{b}}\quad\text{subject to}\ \{\eqref{eq:BSPower},\eqref{eq:AntennaPower},\eqref{eq:SWEE:PowerCons:epi},\eqref{eq:SWEE:rate:epi},\eqref{eq:approx:1-1},\eqref{eq:approx:2-1}\}.\label{Prob:PWEE:approx}
\end{align}

\subsubsection{Max-Min Fairness Energy Efficiency}

The problem of maxminEE is given by
\begin{equation}
\underset{{\bf v}}{\text{maximize}}\enskip\underset{b\in{\cal B}}{\text{min}}\frac{\sum_{u\in{\cal {U}}_{b}}r_{b_{u}}({\bf v})}{P_{\text{BS,}b}({\bf v})}\quad\text{subject to}\enskip\bigl\{\eqref{eq:BSPower},\eqref{eq:AntennaPower}\bigr\}.\label{Prob:maxminEE}
\end{equation}
We note that the centralized and distributed solutions for \eqref{Prob:maxminEE}
have been provided in \cite{Giang:15:JCOML} and \cite{Giang:2017:WCOM},
respectively. Here for complete discussion and self containment, we
provide the main steps of solving the problem. Specifically, with
the newly introduced variables $\eta,\{z_{b}\},\{t_{b}\},\{g_{b_{u}}\}$,
\eqref{Prob:maxminEE} can be equivalently written as
\begin{equation}
\underset{{\bf v},\eta,\{z_{b}\},\{t_{b}\},\{g_{b_{u}}\}}{\text{maximize}}\enskip\eta\quad\text{subject to}\enskip\bigl\{\eta\leq\frac{z_{b}^{2}}{t_{b}},\ \forall b\in{\cal B};\eqref{eq:BSPower},\eqref{eq:AntennaPower},\eqref{eq:SWEE:PowerCons:epi},\eqref{eq:SWEE:rate:epi},\eqref{eq:SWEE:SINR:epi}\bigr\}\label{Prob:maxminEE:epi}
\end{equation}
where $\eta$ represents the minimum EE among all BSs. A convex approximation
of \eqref{Prob:maxminEE:epi} solved in iteration $n$ is
\begin{align}
\underset{{\bf v},\eta,\{z_{b}\},\{t_{b}\},\{g_{b_{u}}\}}{\text{maximize}} & \enskip\eta\quad\text{subject to}\ \{\eqref{eq:BSPower},\eqref{eq:AntennaPower},\eqref{eq:SWEE:PowerCons:epi},\eqref{eq:SWEE:rate:epi},\eqref{eq:approx:1-1},\eqref{eq:approx:2-1}\}.\label{Prob:maxminEE:approx}
\end{align}

\subsection*{SOCP Formulations of Approximate Programs\label{subsec:SOCP-Formulations}}

It is clear that the convex approximate problems \eqref{Prob:NEE:approx},
\eqref{Prob:SWEE:approx}, \eqref{Prob:PWEE:approx} and \eqref{Prob:maxminEE:approx}
are general convex programs due to the logarithmic constraints, i.e.,
\eqref{eq:NEE:rate:epi} and \eqref{eq:SWEE:rate:epi}, and the nonlinear
model of PA's efficiency in \eqref{eq:NEE:PowerCons:epi} and \eqref{eq:SWEE:PowerCons:epi}.
Although off-the-shelf solvers are applicable to solve such programs,
the computational complexity to output solutions is relatively high
in general \cite{BenNemi:book:LectModConv}. Interestingly, it turns
out that these constraints can be represented by second-order-cone
(SOC) constraints which can take the advantages of more powerful SOCP-solvers
to reduce the computational effort. In the rest of this section, we
discuss methods that can invoke the hidden SOC-representation of the
approximated convex programs.

We first consider constraint \eqref{eq:SWEE:PowerCons:epi} which
can be equivalently transformed as
\begin{align}
\eqref{eq:SWEE:PowerCons:epi}\Leftrightarrow\begin{cases}
\sqrt{\sum_{u\in{\cal {U}}_{b}}|[{\bf v}_{b_{u}}]_{m}|^{2}}\leq u_{b,m},\forall b\in{\cal B},m=1,...,M\\
t_{b}\geq\sum_{m=1}^{M}u_{b,m}+P_{\text{cir},b},\forall b\in{\cal B}
\end{cases}\label{eq:antenna:SOCP:epi}
\end{align}
where $\{u_{b,m}\}_{b,m}$ are slack variables. We can see that the
first type of constraint in the equivalent formulation is SOC while
the second one is linear. Constraint \eqref{eq:NEE:PowerCons:epi}
is treated similarly and skipped for conciseness.

We now focus on constraint \eqref{eq:SWEE:rate:epi} whose equivalent
formulation is given as
\begin{align}
\eqref{eq:SWEE:rate:epi}\Leftrightarrow\begin{cases}
z_{b}^{2}\leq\sum_{u\in \mathcal{U}_{b}}\beta_{b_{u}},\forall b\in{\cal B}\\
\log(1+g_{b_{u}})\geq\beta_{b_{u}},\forall b_{u}
\end{cases}\label{eq:SOCP:rate}
\end{align}
where $\{\beta_{b_{u}}\}_{b_{u}}$ are slack variables. Remark that
\eqref{eq:SOCP:rate} is SOC-representable \cite[Sect. 3.3]{BenNemi:book:LectModConv}.
Because the first type of constraint on the right side is SOC-representable,
we only have to deal with the second one. From now on, for notational
convenience, we consider constraint $\log(1+x)\geq y$ where $x,y$
are positive variables. In the following, we provide four different
approaches translating the constraint into SOC-representations.

\subsubsection*{Conic Approximation of Exponential Cone}

The first approach approximates $\log(1+x)\geq y$
by a set of conic constraints based on the result in \cite[Example 4]{BenNemi:01:Onthepolyhedral},
which has been particularly applied to reduce complexity of solving
EE problems in \cite[(31)]{Oskari:EE-optimalBeamDesign:14:JSP} and
\cite[(13)]{Giang:15:JCOML}. The detailed formulation of the conic constraints approximating
logarithmic function is omitted here due to the space limitation.

In some settings, using conic approximation of exponential
cone could cause a significant increase in per-iteration complexity
due to a large number of additional slack variables. This issue is
avoided by the approaches presented following.

\subsubsection*{Equivalently SCA-applicable Constraint}

The second approach equivalently rewrites $\log(1+x)\geq y$ as a
nonconvex but SCA-applicable constraint. To see this, let us multiply
both sides of the constraint by $x$, i.e.
\begin{align}
x\log(1+x)\geq xy\label{eq:SR:epi:1}
\end{align}
Since $x\log(1+x)$ is convex, we can apply SCA principles on \eqref{eq:SR:epi:1}.
A lower bound of $x\log(1+x)$ is given as
\begin{equation}
x\log(1+x)\geq d^{(n)}x-c^{(n)}\label{eq:logapp-conic}
\end{equation}
where $c^{(n)}\triangleq\frac{(x^{(n)})^{2}}{x^{(n)}+1}$, $d^{(n)}\triangleq\frac{x^{(n)}}{x^{(n)}+1}+\log(1+x^{(n)})$,
and $x^{(n)}$ is some positive value. Then, an approximation of \eqref{eq:SR:epi:1}
is
\begin{equation}
d^{(n)}x-c^{(n)}\geq xy\label{eq:logapp}
\end{equation}
which can be represented as SOC constraint, i.e,
\begin{equation}
\|[x+y-d^{(n)}\hspace{1em}2\sqrt{c^{(n)}}]\|_{2}\leq x-y+d^{(n)}.\label{eq:conic:logbound:1}
\end{equation}

\subsubsection*{Concave Lower Bound of the Logarithm}

We can use the well-known inequality of logarithmic function given
as
\begin{equation}
\log(1+z)\geq z(1+z)^{-1}\label{eq:log:ineq}
\end{equation}
for all $z>-1$. By replacing $z$ on both sides of \eqref{eq:log:ineq}
by $\frac{x-x^{(n)}}{x^{(n)}+1}$ we arrive at
\begin{equation}
\log(1+x)\geq\log(1+x^{(n)})+(x-x^{(n)})(1+x)^{-1}\label{eq:log:approx}
\end{equation}
for all $x\geq0$. Now, we can easily check that \eqref{eq:log:approx}
satisfies three conditions (a)\textendash (c). Thus the valid approximate
of $\log(1+x)\geq y$ is
\begin{equation}
\log(1+x^{(n)})+(x-x^{(n)})(1+x)^{-1}\geq y.\label{eq:rate:approx}
\end{equation}
Interestingly, \eqref{eq:rate:approx} contains a hidden SOCP representation
given as
\begin{gather}
\|2\sqrt{1+x^{(n)}},\ \log(1+x^{(n)})-y-x\|_{2}\leq\log(1+x^{(n)})-y+x+2.\label{eq:conic:logbound:2}
\end{gather}

\subsubsection*{Quadratic Lower-bound of the Logarithm}

We can directly approximate \eqref{eq:SWEE:rate:epi} under SCA principles
without requiring the transformation step \eqref{eq:SOCP:rate}. Specifically,
we use the following concave quadratic lower-bound derived based on
the Lipschitz continuity of the logarithm \cite{Oskari:2017:TSP}
\begin{equation}
\log(1+x)\geq\log(1+x^{(n)})+\frac{(x-x^{(n)})}{1+x^{(n)}}-\frac{C}{2}(x-x^{(n)})^{2}.\label{eq:rate:approx:Lipschitz}
\end{equation}
With $C\geq1$, the inequality holds for all $x\geq0$ and $x^{(n)}\geq0$.
As a result, an approximation of \eqref{eq:SWEE:rate:epi}
can be written as
\begin{equation}
z_{b}^{2}+\frac{C}{2}(g_{b_{u}}-g_{b_{u}}^{(n)})^{2}\leq\sum_{u\in \mathcal{U}_{b}}\log(1+g_{b_{u}}^{(n)})+\frac{(g_{b_{u}}-g_{b_{u}}^{(n)})}{1+g_{b_{u}}^{(n)}}\label{eq:conic:Lipschitz}
\end{equation}
which is indeed a rotated-SOC constraint. It is worth noting that
constant $C$ has large impact on the tightness of the approximation
\eqref{eq:rate:approx:Lipschitz}, and thus, it influences the convergence
speed of the iterative algorithm. More specifically, a smaller value
of $C$ implies a tighter approximation, and may increase the convergence
speed (the discussion is numerically justified in Fig. \ref{Fig.6}).

\subsection{Distributed Implementation}

The algorithms in Subsection \ref{subsec:SCA-based-Solutions} are
designed in a centralized fashion under the assumption that each BS (or a central controller)
perfectly knows all channel state information in the network. From
the practical perspective, distributed solutions may be more attractive.
Note that the conventional approaches are not suitable for decentralized
implementation, since updating the parameterized values requires a
central node \cite{DoanhTCOM,Doanh:15:SPL:EEforMultiHoming}. In contrast,
the SCA-based algorithms can be easily carried out in distributed
manner. In fact, distributed implementations of the SCA solutions
for EEmax problems have been provided in \cite{Oskari:DecentralizedMISO,Giang:ICC:2016,Oskari:2017:TSP}.
We remark that distributed implementation is preferred to SWEE, WPEE
and EE-fairness due to their goal of achieving EE of individual node.
In what follows, we present how to solve SWEE maximization problem
\eqref{Prob:SWEE} distributively, and note that the procedure can
be applied to WPEE and EE-fairness problems with slight modifications.

We first assume that each BS has (perfect) CSIs of the channels from
itself to all users in the network, which is referred to as local
CSI. This is a basic assumption in the distributed setting which has
been adopted in \cite{Oskari:2017:TSP,Oskari:DecentralizedMISO,Giang:ICC:2016}.
The main idea of the proposed approach is to solve the convex subproblem
\eqref{Prob:SWEE:approx} distributively using the alternating direction
method of multipliers (ADMM) \cite{Boyd:11:ADMM}. To do so, the vital
step is to recognize the terms which need to be decoupled. From \eqref{Prob:SWEE:approx},
we observe that excluding the inter-cell interference terms $\{G_{b_{u}}(\mathbf{v})\}$,
all the other terms are readily local. For clarity, let us rewrite
\eqref{eq:approx:2-1} as
\begin{equation}
\frac{2\Re(\tilde{\mathbf{h}}_{b,b_{u}}^{(n)}\mathbf{v}_{b_{u}})}{q_{b_{u}}^{(n)}}-\frac{|\mathbf{h}_{b,b_{u}}\mathbf{v}_{b_{u}}^{(n)}|{}^{2}q_{b_{u}}}{(q_{b_{u}}^{(n)})^{2}}\geq\bar{G}_{b_{u}}(\bar{\mathbf{v}}_{b})+\sum_{k\in{\cal B}\backslash\{b\}}\sum_{i\in{\cal U}_{k}}|{\bf h}_{k,b_{u}}{\bf v}_{k_{i}}|^{2},\ \forall b\in{\cal B},u\in{\cal U}_{b}\label{eq:approx:2-1:rewrite}
\end{equation}
where $\bar{\mathbf{v}}_{b}\triangleq[{\bf v}_{b_{1}}^{T},{\bf v}_{b_{2}}^{T},...,{\bf v}_{b_{U}}^{T}]^{T}$,
and $\bar{G}_{b_{u}}(\bar{\mathbf{v}}_{b})\triangleq\sum_{i\in{\cal U}_{b}\backslash\{{u}\}}|{\bf h}_{b,b_{u}}{\bf v}_{b_{i}}|^{2}+\sigma_{b_{u}}^{2}$represents
the intra-cell interference plus noise. To deal with the inter-cell
interference, let us introduce variables $\{\theta_{k,b_{u}}\}_{k,b_{u}}$.
Then \eqref{eq:approx:2-1:rewrite} is equivalent to the following
set of constraint

\begin{align}
\frac{2\Re(\tilde{\mathbf{h}}_{b,b_{u}}^{(n)}\mathbf{v}_{b_{u}})}{q_{b_{u}}^{(n)}}-\frac{|\mathbf{h}_{b,b_{u}}\mathbf{v}_{b_{u}}^{(n)}|{}^{2}q_{b_{u}}}{(q_{b_{u}}^{(n)})^{2}}\geq\bar{G}_{b_{u}}(\bar{\mathbf{v}}_{b})+\sum_{k\in{\cal B}\backslash\{b\}}\theta_{k,b_{u}},\ \forall b\in{\cal B},u\in{\cal U}_{b}\label{eq:distrans1}\\
\theta_{k,b_{u}}\geq\sum_{i\in{\cal U}_{k}}|{\bf h}_{k,b_{u}}{\bf v}_{k_{i}}|^{2},\ \forall b\in{\cal B},u\in{\cal U}_{b},k\in{\cal B}\backslash\{b\}.\label{eq:distran2}
\end{align}
With the transformation, we turn to handling the term $\sum_{k\in{\cal B}\backslash\{b\}}\theta_{k,b_{u}}$
in \eqref{eq:distrans1} for distributed implementation since the
constraints in \eqref{eq:distran2} can be treated locally. To this
end, let us introduce new local variables for each interference term
$\theta_{b,k_{i}}$ as $\tilde{\theta}_{b,k_{i}}^{b}$ and
$\tilde{\theta}_{b,k_{i}}^{k}$optimized at BS $b$ and BS $k$, respectively.
To make sure that these local variables are equal to each other, we
further add an equality constraint
\begin{align}
\theta_{b,k_{i}}=\tilde{\theta}_{b,k_{i}}^{b}=\tilde{\theta}_{b,k_{i}}^{k},\forall b\in\mathcal{B},k\in\mathcal{B}\setminus\{b\},i\in{\cal U}_{k}.\label{}
\end{align}
Now, we can write \eqref{Prob:SWEE:approx} equivalently as \begin{subequations}\label{Prob:SWEE:consensus}
\begin{align}
\text{maximize} & \quad\sum_{b\in{\cal B}}\omega_{b}\eta_{b}\\
 & \quad\text{s.t}\ (\bar{{\bf v}}_{b},\{\eta_{b}\},\{z_{b}\},\{t_{b}\},\{g_{b_{u}}\},\{\tilde{\boldsymbol{\theta}}_{b}\})\in{\mathcal{S}_{b}},\forall b\in\mathcal{B}\\
 & \quad\boldsymbol{\theta}_{b}=\tilde{\boldsymbol{\theta}}_{b},\forall b\in\mathcal{B}
\end{align}
\end{subequations} where $\boldsymbol{\theta}_{b}\triangleq\{\{\theta_{k,b_{k}}\}_{k\in\mathcal{B}\setminus\{b\},u\in\mathcal{U}_{b}},\{\theta_{b,k_{i}}\}_{k_{i}\in\bar{\mathcal{U}}_{b}}\}$,
$\tilde{\boldsymbol{\theta}}_{b}\triangleq\{\{\tilde{\theta}_{k,b_{u}}^{b}\}_{k\in\mathcal{B}\setminus\{b\},u\in\mathcal{U}_{b}},\{\tilde{\theta}_{b,k_{i}}^{b}\}_{k_{i}\in\bar{\mathcal{U}}_{b}}\}$,
$\bar{\mathcal{U}}_{b}\triangleq\underset{k\in\mathcal{B}\setminus\{b\}}{\cup}{\cal U}_{k}$,
and
\begin{eqnarray} \label{distributed set}
\mathcal{S}_{b} & \triangleq & \Bigl\{(\bar{{\bf v}}_{b},\{\eta_{b}\},\{z_{b}\},\{t_{b}\},\{g_{b_{u}}\},\{\tilde{\boldsymbol{\theta}}_{b}\})\bigr|\nonumber \\
 &  & \quad\quad\frac{2\Re(\tilde{\mathbf{h}}_{b,b_{u}}^{(n)}\mathbf{v}_{b_{u}})}{q_{b_{u}}^{(n)}}-\frac{|\mathbf{h}_{b,b_{u}}\mathbf{v}_{b_{u}}^{(n)}|{}^{2}q_{b_{u}}}{(q_{b_{u}}^{(n)})^{2}}\geq\bar{G}_{b_{u}}(\bar{\mathbf{v}}_{b})+\sum_{k\in{\cal B}\backslash\{b\}}\theta_{k,b_{u}}^{b},\ \forall u\in{\cal U}_{b}\nonumber \\
 &  & \quad\quad\quad\quad\theta_{b,k_{i}}^{b}\geq\sum_{u\in{\cal U}_{b}}|{\bf h}_{b,k_{i}}{\bf v}_{b_{u}}|^{2},\ \forall k_{i}\in{\bar{{\cal U}}_{b}}\nonumber \\
 &  & \quad\quad\quad\quad\quad\quad{\{\eqref{eq:BSPower},\eqref{eq:AntennaPower},\eqref{eq:SWEE:PowerCons:epi},\eqref{eq:SWEE:rate:epi},\eqref{eq:approx:1-1},\eqref{eq:approx:2-1}\}}_{b}\Bigl\}
\end{eqnarray}
is the feasible set for the local variables at BS $b$. In \eqref{distributed set}, the notation
$\{\}_{b}$ means to take only the constraints related to BS $b$.
Problem \eqref{Prob:SWEE:consensus} is in a form of global consensus
problem and can be optimally solved using the standard ADMM procedure.
We refer interested readers to \cite{Oskari:DecentralizedMISO,Giang:2017:WCOM}
for further details.

\section{Numerical result\label{sec:Numerical-result}}

We evaluate the performances of the different algorithms presented
above. The general (fixed) simulation parameters are taken from Table
\ref{Tab. 1} and the ones which are changed in the simulations are
given in the caption of the corresponding figures. A network of $B$
base stations is considered, and $U_{b}$ users are randomly dropped
to the coverage area of each base station. The user channels follow
the Rayleigh distribution.

\begin{table}[tb]
\caption{Simulation Parameters}

\centering{}%
\begin{tabular}{c|c}
\hline
{Parameters}  & {Value}\tabularnewline
\hline
\hline
Path loss and shadowing  & $38\log_{10}\left(d\text{ [m]}\right)+34.5$+$\mathcal{N}(0,8)$\tabularnewline
Inter-BS distance  & $D=1$ km\tabularnewline
Static power consumption $P_{\text{sta}}$  & 33 dBm\tabularnewline
Static power consumption $P_{\text{dyn}}$  & 30 dBm\tabularnewline
Power amplifier efficiency $\epsilon$  & 0.35\tabularnewline
Number of BSs $B$  & 3\tabularnewline
Number of users per cell $U_{b}$  & 2\tabularnewline
Number of Tx antennas $M$  & 4\tabularnewline
Signal bandwidth $W$  & 10 kHz\tabularnewline
Power spectral density of noise  & -174 dBm/Hz\tabularnewline
\hline
\end{tabular}\vspace{-2mm}
 \label{Tab. 1}
\end{table}

\subsection{Comparison on the Convergence and the Performance}

In the first set of simulations, we compare the SCA methods against
the conventional based on the fractional program (FP) ones in terms
of the convergence rate and achieved EE performances. The following
beamforming designs are considered in the evaluation:
\begin{itemize}
\item NEE-SCA: the SCA procedure for solving \eqref{Prob:NEE}.
\item SWEE-SCA: the SCA procedure for solving \eqref{Prob:SWEE}. We set
equal priorities for all the nodes, i.e, $\omega_{b}=1,\forall b$.
\item WPSEE-SCA: the SCA procedure for solving \eqref{Prob:PWEE}. Similar
to SWEE-SCA, $\omega_{b}=1,\forall b$ .
\item maxminEE-SCA: the SCA procedure for solving \eqref{Prob:maxminEE}.
\item NEE-FP: the beamforming design based on FP proposed in \cite{HHJY:13:JCOM}.
\item SWEE-FP: the beamforming design based on FP proposed in \cite{She:2014:WSEE}.
This scheme has been studied for a MIMO channel. However, it can be
easily simplified for MISO channels by setting the number of receive
antenna to one.
\item maxmin-FP: the beamforming design based on FP proposed in \cite{HHJYL:13:JCOML}.
Although this approach has been proposed for a multi-cell joint transmission
system, we can easily simplify it for the multi-cell coordinated beamforming
case by properly rewriting the signal and interference terms \cite[Remark 2]{HHJYL:13:JCOML}.
\end{itemize}
We note that for the WPEE metric, only the performance of the SCA-based
method is studied as beamforming designs for this metric based on
the FP framework have not yet been proposed to the best of our knowledge.
In addition, to reduce the computational burden for the simulations,
we terminate the iterative processes of all the considered algorithms
either when the increase in the objective between two consecutive
iteration is less than $10^{-5}$ or after $10^{6}$ iterations. Also,
for a fair comparison, we only consider the conventional power model
as in \cite{zappone2015energy,DerrickKwanNg:2012:JWCOM:EE_OFDM,She:2014:WSEE,HHJY:13:JCOM,HHJYL:13:JCOML},
i.e. fixed signal processing power and PAs' efficiency. The results
for the general power consumption model are reported in Section \ref{subsec:sim:genPowMod}.

\subsection*{Convergence Comparison of the SCA and PF Algorithms}


\begin{figure}
\subfigure[Convergence rate for one channel realization.]{\includegraphics[width=0.47\columnwidth]{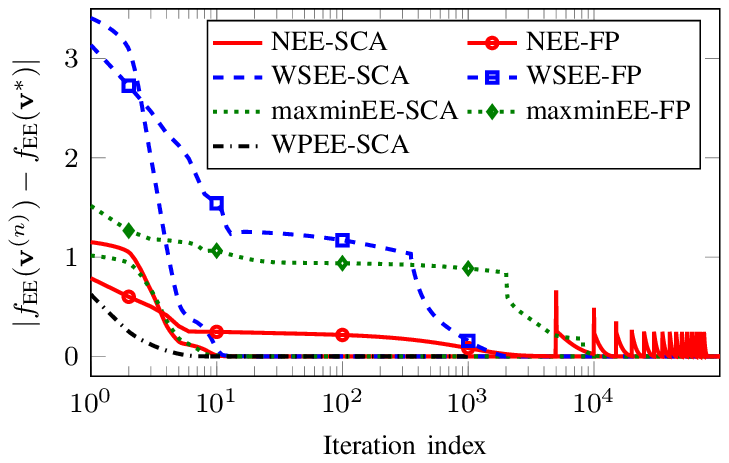}}\enspace{}\subfigure[CDF of the numbers of iterations over 500 channel realizations.]{\includegraphics[width=0.48\columnwidth]{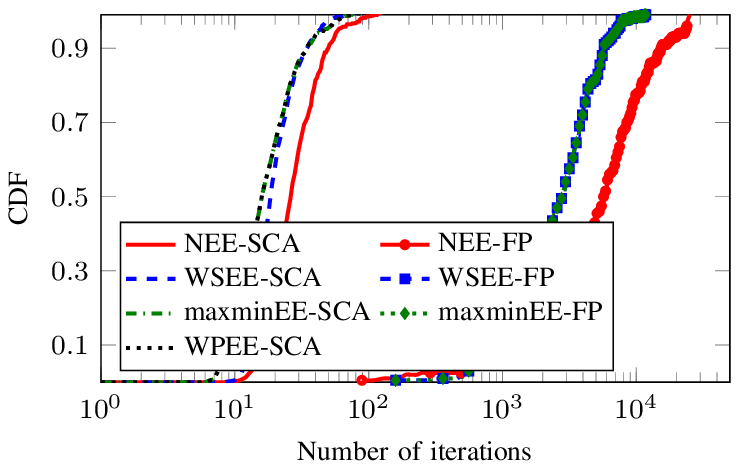}}\caption{Convergence behavior of different EE schemes with $P_{b}=30$ dBm}
\label{Fig.1}
\end{figure}


Fig. \ref{Fig.1} presents the convergence behavior
of the considered EE approaches. Particularly, Fig. \ref{Fig.1}(a)
compares the convergence speed of the SCA frameworks and the FP methods
in terms of the number of iterations for one random channel realization
by showing the gap between the current objective value at iteration
$n$ and the achieved objective value after the termination of the
algorithms, i.e., $|f_{\text{EE}}({\bf v}^{n})-f_{\text{EE}}({\bf v}^{\ast})|$.
As can be seen, the SCA based methods have steady monotonic converge
properties, e.g., achieve the objective value of the convergence point
after ten iterations in the considered setting. For the FP based approaches,
even hundreds or thousands of iterations can be required to reach
the convergence, while the monotonic convergence is not always guaranteed,
e.g., for the NEE-FP method.

To complete the comparison in terms of convergence speed between the
SCA and FP-based methods, we provide in Fig. \ref{Fig.1}(b) the cumulative
distribution function (CDF) of the total number of iterations needed
for convergence. It is observed that for 90\% of channel realizations,
the SCA converges after 30 iterations while the FP methods need even
thousands of iterations to terminate. This observation again shows
the superiority of the SCA algorithms in terms of complexity compared
to the conventional approaches.

\subsection*{EE Performance Comparison of the SCA and FP Algorithms}

\begin{figure}
\centering{}\subfigure[Network EE.]{\includegraphics[width=0.465\columnwidth]{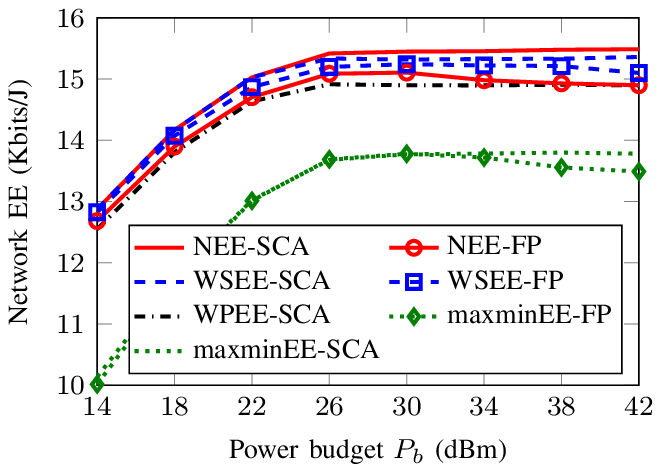}}\quad \subfigure[Sum EE.]{\includegraphics[width=0.45\columnwidth]{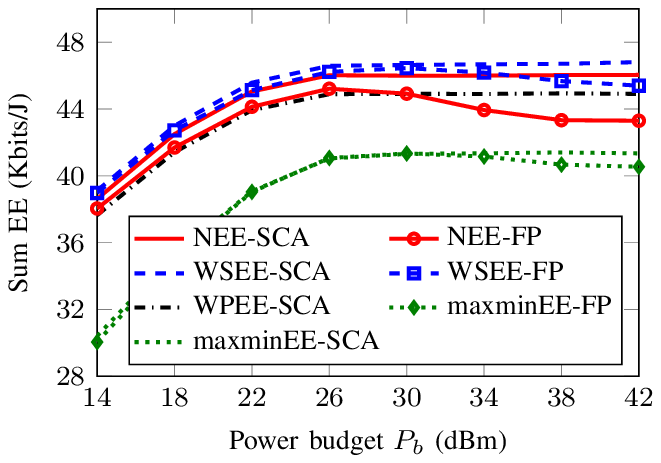}}\\
\smallskip{}
\subfigure[Minimum EE.]{\includegraphics[width=0.45\columnwidth]{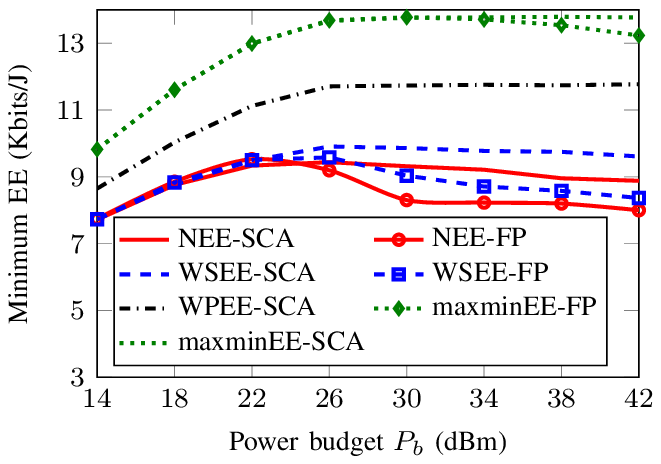}}\caption{Achieved EE performance versus $P_{b}$}
\label{Fig.3}
\end{figure}

Fig. \ref{Fig.3} plots the average performances of the SCA and FP
based methods in terms of the achieved NEE, the sum EE and the minimum
EE versus the maximum transmit power budget $P_{b}$. Our first observation
is that the SCA method maximizing a specific EE metric achieves approximately
the same EE performance compared to the corresponding FP method in
small and moderate power regions. This again implies the effectiveness
of the SCA framework in solving the EE maximization problems as it
can offer similar performance compared to the conventional ones
but with much reduced complexity. Another observation is that the
achieved EE with all the approaches saturates when the power budget
is sufficiently large. This is because in the large power regime,
the data rate logarithmically scales with the transmit power while
the power consumption increases linearly with the transmit power.
Thus, whenever the gain in achieved throughput cannot compensate for
the increase of power consumption, the EE methods do not use the excess
power to further increase data rate so as to maintain a high value EE. This
fact has been discussed in many EE maximization related works \cite{zappone2015energy,DerrickKwanNg:2012:JWCOM:EE_OFDM,Dan:2013:SP:Fullduplex}.
On the other hand, we can see that the EEs achieved by the FP methods
slightly downgrade for large value of $P_{b}$. The reason can be
explained as follows. When $P_{b}$ increases, the feasible set of
the EE problems is expanded which results in the increasing number
of iterations required for the convergence. However, due to the threshold
on the maximum iterations for the iterative process, the FP methods
may not reach the suboptimal solutions within $10^{6}$ iterations.
Consequently, they may output poor performances leading to the decrease
of average achieved EE value. This observation again points out the
drawback of the two-layer iterative procedure in practice. Let us
then evaluate the achieved performances of the considered methods
with respect to each EE metric. It is obvious that the NEE maximization
methods outperform the other schemes in terms of the achieved NEE
(in Fig. \ref{Fig.3}(a)), while the SWEE methods offer the best sum-EE
values (in Fig. \ref{Fig.3}(b)). In terms of minimum EE, the maxminEE methods achieve the best performance as they aim at maintaining the balance of EE among all parties (in Fig. \ref{Fig.3}(c)). However, the maxminEE approaches suffer a loss in NEE
and sum-EE performances. The WPEE metric, as expected, offers a better
minimum EE than the SWEE and NEE criteria. We note that individual
EE is one of the key features in many network scenarios (e.g. heterogeneous
networks), and, thus, a per-node EE performance and EE fairness will
be the main focus in the next numerical experiment.

\subsection{Achieved per-BS EE Performance}




\begin{figure}
\centering{}\subfigure[Sum EE.]{\includegraphics[width=0.45\columnwidth]{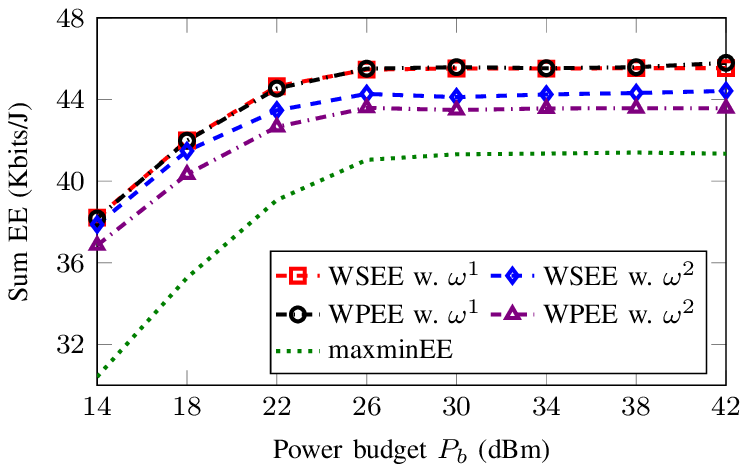}}\enskip{}\subfigure[Maximum EE.]{\includegraphics[width=0.45\columnwidth]{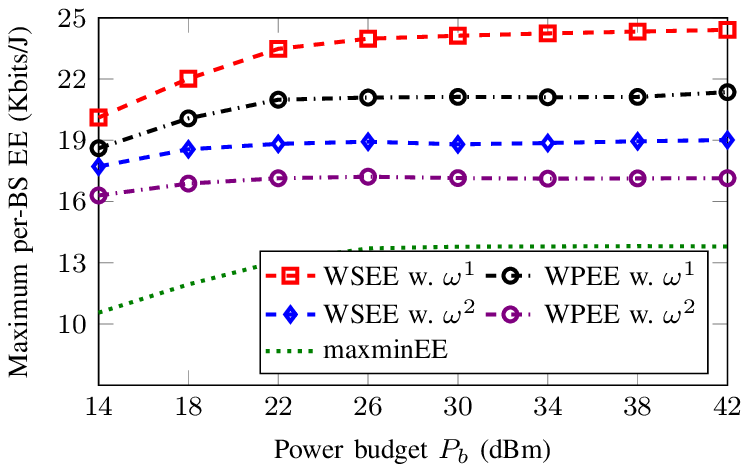}}\\
\smallskip{}
\subfigure[Minimum EE.]{\includegraphics[width=0.45\columnwidth]{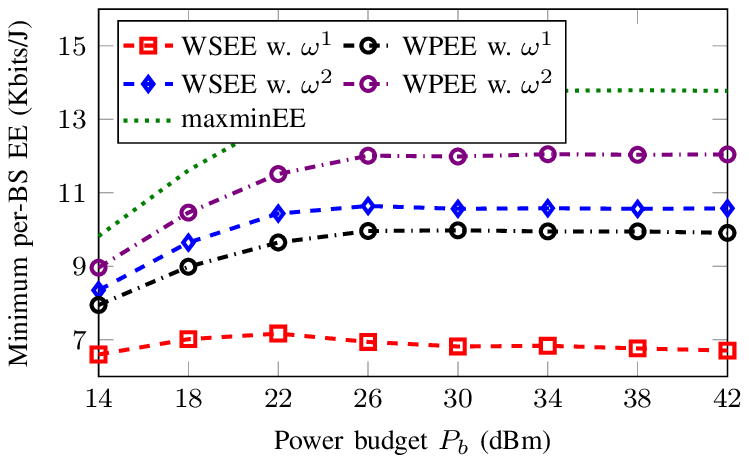}}\enskip{}\subfigure[Fairness index.]{\includegraphics[width=0.45\columnwidth]{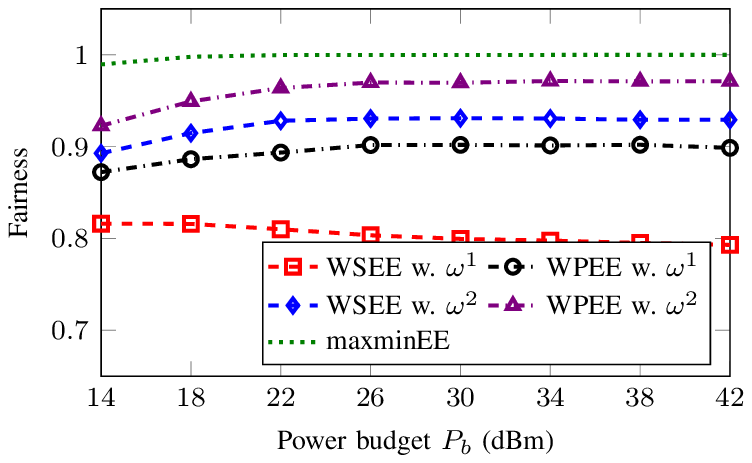}}\caption{Averaged EE versus transmit power budget $P_{b}$. For WSEE and WPEE
schemes, we use two sets of priority parameters, i.e., $\bm{\omega}^{1}=[0.7,0.5,0.3]$
and $\bm{\omega}^{2}=[0.3,0.5,0.7]$.}
\label{Fig.4}
\end{figure}

In Fig. \ref{Fig.4}, we compare the achieved fairness when using different fairness-oriented EE metrics, i.e.,
SWEE, WPEE and maxminEE as a function of the maximum transmit power
budget $P_{b}$. Note that the fairness is considered in terms of energy efficiency, and not rates. Specifically, we consider two settings of weighting
vector i.e., $\bm{\omega}^{1}=[0.7,0.5,0.3]$ and $\bm{\omega}^{2}=[0.3,0.5,0.7]$,
where each value implies the priority weight for a corresponding BS. In this experiment, users in cell 1 are dropped in the radius
of 200 meters to its serving BS while for cell 3, users' locations
are near the cell edge. The served users of BS 2 are randomly placed
in its coverage region. The figure is plotted to see how the priority
parameters alter the per-BS EE behaviors. The average sum-EE, maximum,
minimum EE and EE fairness measure\footnote{The EE fairness measure is calculated following the
index defined in \cite[ (1)]{jain1984quantitative}. Particularly,
let us denote $\bm{\eta}^{\ast}=[\eta_{1}^{\ast},\ldots\eta_{B}^{\ast}]$
as the achieved per-BS EEs after solving, e.g., WSEE problems (28), WPEE problem (33) or maxminEE (36). The fairness index is $\text{fairness}=\frac{\bigl(\sum_{b=1}^{B}\eta_{b}^{\ast}\bigr)^{2}}{B\sum_{b=1}^{B}(\eta_{b}^{\ast})^{2}}$.} among all the BSs are plotted in Figs. \ref{Fig.4}(a)
\textendash \ref{Fig.4}(d), respectively. Similar results as in Fig.
\ref{Fig.3} can be observed. Particularly, the achieved EE values
remain unchanged when $P_{b}$ is sufficiently large. Moreover, it
is seen that the SWEE methods outperform the other schemes in terms
of sum-EE and maximum per-BS EE values. This is clear since maximizing
the sum of individual EEs is the objective of the SWEE methods. In
terms of minimum EE among all nodes, it is obvious that the maxminEE
scheme obtains the best performance followed by the WPEE and SWEE
criteria (with same assigned priority). Another important observation
is that by assigning different priority weights $\bm{\omega}$ for
SWEE and WPEE metrics, we can adjust the achieved EE of each node.
It is discovered that with $\bm{\omega}^{1}$, more priority is given
to BS 1 leading to an improvement in the sum-EE performance for these
two schemes. In addition, since BS 3 is more penalized, the gap between
maximum and minimum per-BS EE values is enlarged and, thus, implying
high EE unfairness among the BSs. On the contrary, since $\bm{\omega}^{2}$
prioritizes BS 3 and restricts BS 1, it reduces the sum-EE performance
of the network but encourages the fairness among all parties. As a
conclusion, the EE fairness measure in Fig. \ref{Fig.4}(d) shows that
the SWEE and WPEE schemes can tune the EE fairness of the system by
the priority parameters while the maxminEE can establish the absolute
fairness among all the per-BS EEs. Also, with the same weighting vector,
the WPEE metric outperforms the SWEE in terms of EE balancing. The
WPEE achieves a better trade-off between fairness and EE performance
compared to the two other schemes SWEE and maxminEE.

\subsection{SCA With Different Conic Approximations}
\begin{figure}
\subfigure[Convergence rate for one random channel realization.]{\label{Fig.6.a}\includegraphics[width=0.45\columnwidth]{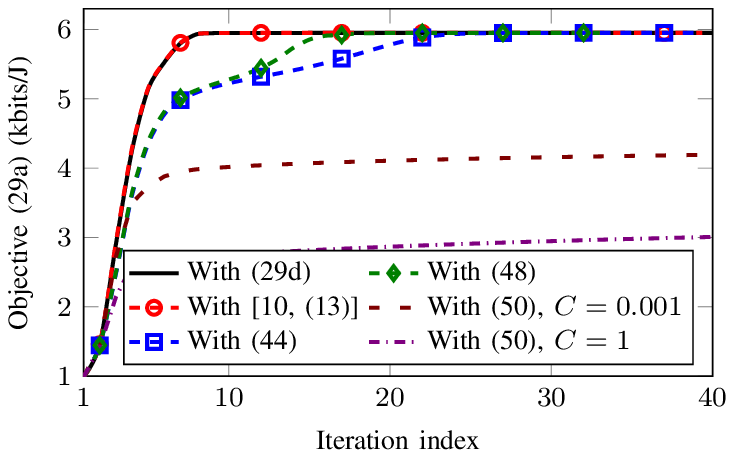}}\enspace{}\subfigure[CDF of the number of iterations over 1000 channel realizations.]{\includegraphics[width=0.45\columnwidth]{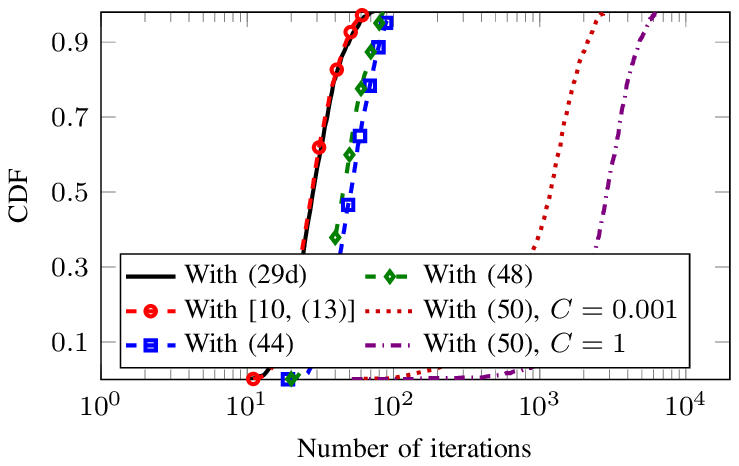}\label{Fig.6.b}}\caption{Convergence behavior for the SCA algorithm solving \eqref{Prob:SWEE:approx}
with different conic approximations with $P_{b}=40$ dBm.}
\label{Fig.6}
\end{figure}
We now illustrate the performances of the SOCP formulations provided
in Section \ref{subsec:SOCP-Formulations}, by focusing on the SWEE
metric. In Fig. \ref{Fig.6}(a), we compare the convergence rate
of the objective \eqref{eq:obj:SWEE:epi} when the original logarithmic
constraint \eqref{eq:SWEE:rate:epi} and its SOC constraints \cite[(13)]{Giang:15:JCOML},
\eqref{eq:conic:logbound:1}, \eqref{eq:conic:logbound:2} and \eqref{eq:conic:Lipschitz}
are used in \eqref{Prob:SWEE:approx}. It is seen that with \cite[(13)]{Giang:15:JCOML},
the algorithm converges with the same rate as using \eqref{eq:SWEE:rate:epi}
and faster than the other SOC approximations. This is understandable
because the set of conic constraints \cite[(13)]{Giang:15:JCOML}
is in fact a tight approximation of logarithmic function up to a fixed
accuracy level. On the other hand, the other ones are the upper bounds
of the logarithmic function which are tight only in the fixed point
at each iteration. Between \eqref{eq:conic:logbound:1} and \eqref{eq:conic:logbound:2},
we can observe that \eqref{eq:conic:logbound:2} offers a better convergence
rate. This may be understood as \eqref{eq:conic:logbound:2} directly
approximates the bound of logarithmic function while \eqref{eq:conic:logbound:2}
is derived from the bound of the equivalent transformation of \eqref{eq:SWEE:rate:epi}.
For the SOC approximation based on \eqref{eq:conic:Lipschitz}, it
is seen that the convergence behavior largely depends on $C$. When
a large value of $C$ is used, the objective slowly converges while
the convergence rate significantly improves when using small value
of $C$. The performance agrees with the analysis of \eqref{eq:conic:Lipschitz}
which argues that the smaller $C$ provides a tighter approximation
in \eqref{eq:conic:Lipschitz} and, thus, can lead to a faster convergence.

In Fig. \ref{Fig.6}(b), we depict the CDF of the number of iterations
required for convergence with different SOCP formulations of \eqref{Prob:SWEE:approx}.
As expected, the result is consistent with that observed from Fig. \ref{Fig.6}(a).
Specifically, adopting the set of conics constraints \cite[(13)]{Giang:15:JCOML}
to approximate the logarithmic function does not require more iterations
for convergence compared to using the original constraint \eqref{eq:SWEE:rate:epi}.
Also, the number of iterations of \eqref{Prob:SWEE:approx} with \eqref{eq:conic:logbound:2}
is smaller than that of applying \eqref{eq:conic:logbound:1}. In
general, we can see that for 90\% of the channel realization, the
approximation methods \cite[(13)]{Giang:15:JCOML},
\eqref{eq:conic:logbound:1}, \eqref{eq:conic:logbound:2} can
provide a good convergence rate which is smaller than 100 iterations.
On the other hand, \eqref{eq:conic:Lipschitz} results in slow convergence
speed in the considered setting.


\subsection{Achieved Performance with General Power Consumption Model\label{subsec:sim:genPowMod}}
\begin{figure}
\begin{minipage}{0.47\columnwidth}\centering\includegraphics[width=1\columnwidth]{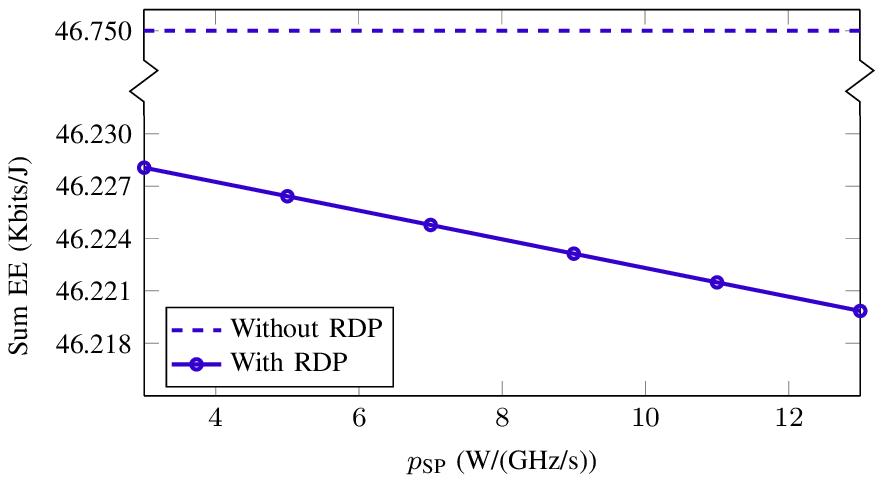}
\caption{Achieved sum EE of WSEE scheme with and without considering rate-dependent
power consumption. We take $P_{b}=40$ dBm. }
\label{Fig.8}\end{minipage}\hfill\begin{minipage}{0.47\columnwidth}\centering\medskip{}
\includegraphics[width=0.9\columnwidth]{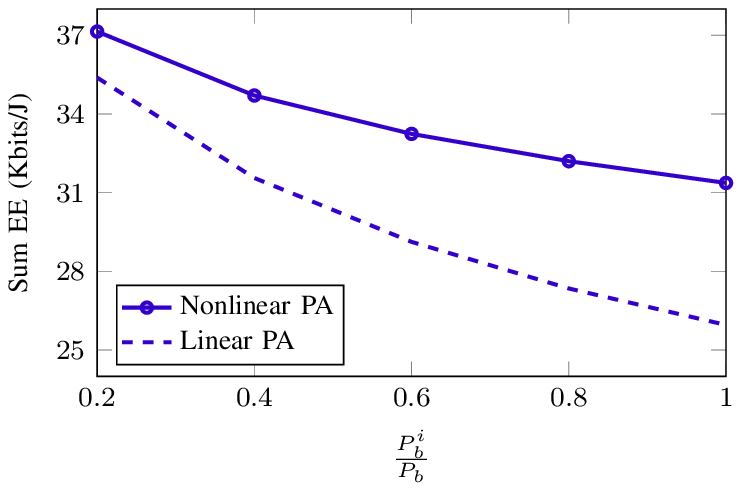}\vspace{-4mm}\caption{Achieved sum EE of WSEE scheme versus ratio $\frac{P_{b}^{m}}{P_{b}}$
for two designs considering linear and nonlinear model of PA's efficiency.
We take $P_{b}=40$ dBm.}
\label{Fig.9}\end{minipage}
\end{figure}
In this numerical experiment, we provide insights to the impact of
rate-dependent signal processing power and the nonlinear model of
PA's efficiency on the achieved EE performance. Fig. \ref{Fig.8}
compares the achieved sum-EE of the SWEE schemes without and with
considering the rate-dependent power (RDP) consumption, which are
labeled as `Without RDP' and `With RDP', respectively. The curve `Without
RDP' is obtained by solving problem \eqref{Prob:SWEE:epi} with $p_{\text{SP}}=0$.
Then, we recalculate the EE values for `With RDP' scheme with the
given $p_{\text{SP}}$ in the horizontal axis. Our observation is
that as $p_{\text{SP}}$ increases, the achieved EE monotonically
decreases. This is understandable because, for a fixed achieved data
rate, higher $p_{\text{SP}}$ increases the total power consumption
and thus, degrades the EE. This result suggests that RDP may be included
when optimizing the EE performance of a wireless network.

Next, in Fig. \ref{Fig.9}, we evaluate the impact of nonlinear PA's efficiency
on the sum-EE achieved by the SWEE method. For this purpose, we perform the
EE optimization based on the nonlinear PA's efficiency model \eqref{eq:nonlinear-PA-Eff}.
As PA's efficiency does not depend on $P_{b}$ but $P_{b}^{m}$, we fix $P_{b}=40$ dBm and plot the achieved sum EE as a function
of the ratio $\frac{P_{b}^{m}}{P_{b}}$. The two following schemes
are compared:
\begin{itemize}
\item `Nonlinear PA': problem \eqref{Prob:SWEE} is solved
using the nonlinear PA's efficiency model \eqref{eq:nonlinear-PA-Eff} with
$\epsilon_{\text{max}}=0.35$.
\item `Linear PA': problem \eqref{Prob:SWEE} is solved
using the PA's power consumption model \eqref{eq:P:PA:linear} with fixed
PA's efficiency $\epsilon=0.35$. The resulting beamforming solution
is used to compute the actual EE performance following the PA's efficiency
model \eqref{eq:nonlinear-PA-Eff}.
\end{itemize}
As can be seen, `Linear PA' scheme is inferior to
`Nonlinear PA' one which clearly shows that the power modeling has
remarkable influence on the achieved EEs. More specifically, the
EE maximization based on the assumption that the PA's efficiency is the same regardless of the output power potentially
degrades the EE performance in practical implementation, where the PA's efficiency actually depends on the desired output power \cite{Mikami2007,Bjoernemo2009,AmplifierMIMO-Persson,Tervo:nonlinearPA}. Another
observation is that the achieved EEs of both schemes decrease against
the increase of $\frac{P_{b}^{m}}{P_{b}}$. The result can be explained
as follows. Recall that the effective PA's efficiency depends on $P_{b}^{m}$
and the actual transmit power (ATP) on the antenna, that is, with
increasing $P_{b}^{m}$ the efficiency slope of that PA is changed so that the
efficiency is worse in the lower ATP regime (see \eqref{eq:nonlinear-PA-Eff}).
Thus, the decreased PA efficiency simply deteriorates the achieved
EE.



\subsection{Achieved EE in Large-scale network settings}


\begin{figure}[t]
\begin{minipage}{0.47\columnwidth}\centering\includegraphics[width=1\columnwidth]{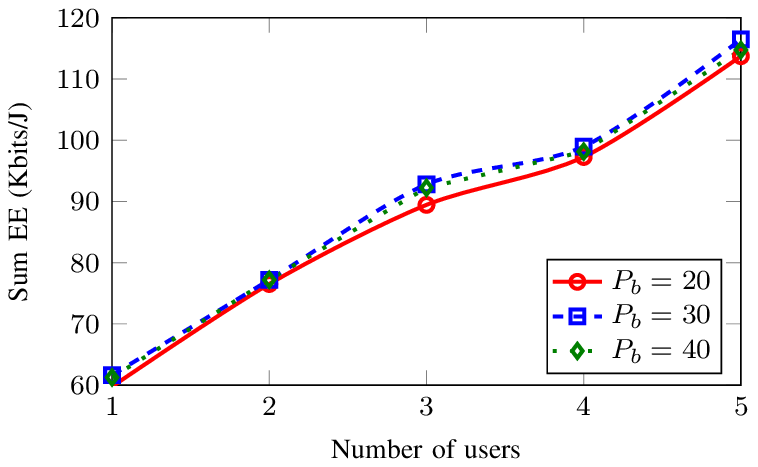}\vspace{-3mm}

\caption{Achieved sum-EE of the SWEE scheme versus the number of users per
cell $U_{b}$ with $B=7$, $M=4$.}
\label{Fig.large1}\end{minipage}\hfill\begin{minipage}{0.47\columnwidth}\centering
\includegraphics[width=0.96\columnwidth]{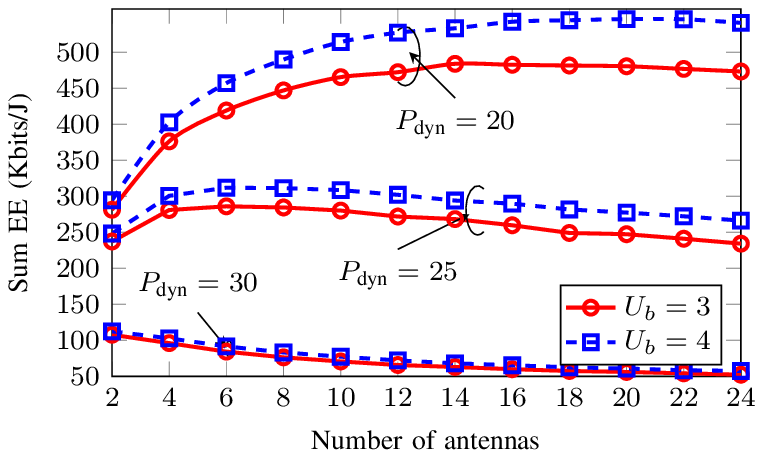}\vspace{-3mm}\caption{Achieved sum-EE of the SWEE scheme versus the number of antennas per
BS with $B=7$. We take $P_{b}=40$ dBm.}
\label{Fig.large2}\end{minipage}
\end{figure}
In the final set of numerical experiments, we illustrate how the achieved
EE behaves in a larger network. A 7-cell network model which consists
of $B=7$ BSs is considered. For simplicity, we adopt the conventional
power consumption model with \eqref{eq:P:PA:linear} and $p_{\text{SP}}=0$
and simulate only the SWEE scheme. Fig.\ \ref{Fig.large1} shows
the achieved EE versus the number of users per cell $U_{b}$ with
different values of transmit power $P_{b}$ when the number of per-BS
antennas is fixed to $M=4$. It is seen that the EE values increase
with the increasing number of served users. This is because the sum
rate is an increasing function of the number of users and, thus, increased
when more users are involved in the transmission. We can also observe
that for fixed $U_{b}$, the EE grows if the power budget is larger.
However, when $P_{b}$ is large enough, further increasing $P_{b}$
does not bring significant improvement in EE. This result is consistent
with that observed in Figs. \ref{Fig.3} and \ref{Fig.4}.

In Fig.\ \ref{Fig.large2}, we show the achieved EE versus the number
of per-BS antennas $M$ for different values of $P_{\text{dyn}}$
and $U_{b}$. As can be seen, for small and moderate values of $P_{\text{dyn}}$,
the achieved EE first increases, then decreases as $M$ keeps increasing,
while for the large value of $P_{\text{dyn}}$, it monotonically decreases.
The reason can be explained as follows. Recall the fact that additional
antennas provide more degree-of-freedom which leads to the improvement
in the achievable data rate. However, since total power consumption
linearly scales with the transmit antennas, adding more antennas consumes
more circuit power. Thus, as long as the the benefit offered from
transmitting with additional antennas is beyond the cost of the power
consumption, the achieved EE increases. Otherwise, increasing the
number of transmit antennas degrades the achieved EE. Another observation,
which agrees with the result in Fig. \ref{Fig.large1}, is that adding
more users improves the achieved EE.

\section{Conclusion\label{sec:conclude}}

We have provided a summary and performance comparison of various algorithms
for the problems of EE optimization in multi-cell multiuser MISO downlink,
under four energy efficiency metrics. We have reviewed and presented
the SCA framework to provide efficient solutions for the energy efficiency
optimization. The algorithms have been numerically evaluated and compared
with different fractional programming solutions for the same problems.
The SCA based algorithms have been shown to outperform the existing
FP ones in terms of convergence speed. This paper can be viewed as
a guideline for the application of the SCA in solving the energy-efficient
beamforming designs in particular, and the nonconvex problems in wireless
communications in general.

The EE optimization will be important for the sustainability
of the future digital society. Several important problems still remain.
For example, acquiring accurate CSI is challenging in practice, and the transmission designs taking into account
the effect of imperfect CSI is an important topic to be explored. The impact of data sharing over (wireless) power and bandwidth
limited backhaul in the CoMP joint processing transmission is an important
topic. The hybrid analog/digital beamforming transceiver architecture
based EE optimization for the evolving millimeter wave wireless communications
is an important item for the evolving 5G system design. The cloud
radio access network (CRAN) architecture with more processing options
either close to the antenna or at computing cloud requires also the
EE based design and analysis. Finally, the power consumption in user
devices is much more difficult to model and control than that in the
base stations or cloud, but constitutes a significant portion of the
overall network power consumption.


\section*{Appendix}

\subsection*{Proof of Lemma \ref{lem:equivalence}:\label{appendix:Proof:Lemma:Equi}}

For proving the lemma, we show that constraints \eqref{eq:NEE:EE:epi}\textendash \eqref{eq:NEE:SINR:epi}
are active at the optimality by the contradiction. Let $({\bf v}^{\ast},\eta^{\ast},z^{\ast},t^{\ast},\{g_{b_{u}}^{\ast}\})$
be an optimal solution of \eqref{Prob:NEE:epi} and suppose that \eqref{eq:NEE:SINR:epi}
is not active at the optimum, i.e., $\frac{|{\bf h}_{b,b_{u}}{\bf v}_{b_{u}}^{\ast}|^{2}}{G_{b_{u}}({\bf v}^{\ast})+\sigma_{b_{u}}^{2}}>g_{b_{u}}^{\ast}$
for some $b_{u}$. Then we can scale down the transmit power for user
$b_{u}$ and achieve a new beamformer $\|\hat{{\bf v}}_{b_{u}}\|_{2}^{2}$
such that $\|\hat{{\bf v}}_{b_{u}}\|_{2}^{2}=\tau\|{\bf v}^{\ast}_{b_{u}}\|_{2}^{2}<\|{\bf v}^{\ast}_{b_{u}}\|_{2}^{2}$
for $\tau\in(0,1)$ while keeping the others remaining unchanged,
i.e $\hat{{\bf v}}_{b_{k}}={\bf v}_{b_{k}}^{\ast}$ for all $b_{k}\neq b_{u}$.
By this way, we can achieve $\frac{|{\bf h}_{b,b_{u}}\hat{{\bf v}}_{b_{u}}|^{2}}{G_{b_{u}}(\hat{{\bf v}})+\sigma_{b_{u}}^{2}}>g_{b_{u}}^{\ast}$
for all $b_{u}$ since interference power at all users has reduced.
In addition, we have $\hat{t}=P_{\text{total}}(\hat{{\bf v}})<t^{\ast}=P_{\text{total}}({\bf v}^{\ast})$.
Consequently, we can find $\hat{\eta}=\frac{(z^{\ast})^{2}}{\hat{t}}>\eta^{\ast}$.
This contrasts to the fact that $({\bf v}^{\ast},\eta^{\ast},z^{\ast},t^{\ast},\{g_{b_{u}}^{\ast}\})$
is the optimal solution. The same spirit is applied to the other constraints.
This completes the proof.



\begin{backmatter}


\section*{Availability of data and material}
Not applicable.

\section*{Competing interests}
  The authors declare that they have no competing interests.
\section*{Funding}

This work was supported in part by the projects ``Wireless Connectivity for Internet of Everything - Energy Efficient Transceiver and System Design (WiConIE)" funded by Academy of Finland under Grant 297803, and ``Flexible Uplink-Downlink Resource Management for Energy and Spectral Efficiency Enhancing in Future Wireless Networks (FURMESFuN)" funded by Academy of Finland under Grant 31089.

\section*{Author's contributions}

All authors have contributed extensively to the work presented in this manuscript. Kien-Giang Nguyen performed the numerical experiments, wrote and revised the manuscript. Oskari Tervo took part in numerical experiments, cowrote, reviewed and revised the manuscript. Quang-Doanh Vu critically reviewed, revised and provided suggestions to improve the presentation of the manuscript. Le-Nam Tran reviewed and revised the manuscript. Markku Juntti supervised the work, reviewed and revised the manuscript. All authors read and approved the manuscript.

 \section*{Acknowledgements}
 Not applicable.


%
%

\bibliographystyle{bmc-mathphys} 
\end{backmatter}
\end{document}